\documentclass[times,twocolumn,final]{elsarticle}
\usepackage{xcolor}
\usepackage{cag}
\usepackage{framed,multirow}

\usepackage{amssymb}
\usepackage{latexsym}

\usepackage{url}
\definecolor{newcolor}{rgb}{.8,.349,.1}

\usepackage{hyperref}
\usepackage{array}% http://ctan.org/pkg/array
\usepackage[switch,pagewise]{lineno} %Required by command \linenumbers below
\usepackage{tikz}
\usepackage{graphicx}
\usepackage{amsfonts, amsmath, amssymb}
\usepackage{amscd}
\usepackage[bottom]{footmisc}
\newcommand{\R}{{\mathbb R}}
\newtheorem{theorem}{Theorem}[section]

\newtheorem{proposition}[theorem]{Proposition}
\newtheorem{definition}[theorem]{Definition}

\def\etal{\textit{et al.}}
\begin{document}

\verso{Preprint Submitted for review}

\begin{frontmatter}

\title{Projection-based Classification of Surfaces for 3D Human Mesh Sequence Retrieval}%
%\author[1]{Anonymous submission}
\author[cristal]{Emery \snm{Pierson}}
%\cortext{Corresponding author: \ead{emery.pierson@univ-lille.fr}}
\cortext[cor1]{Corresponding author: }\ead{emery.pierson@univ-lille.fr}

\author[painleve]{Juan-Carlos \snm{\'Alvarez Paiva}}
\author[imt, cristal2]{Mohamed \snm{Daoudi}}
\address[cristal]{Univ. Lille, CNRS, Centrale Lille, UMR 9189 CRIStAL, Lille, F-59000, France}
\address[imt]{IMT Nord Europe Institut Mines-Télécom, Univ. Lille, Centre for Digital Systems, Lille, F-59000, France}
\address[cristal2]{Univ. Lille, CNRS, Centrale Lille, Institut Mines-Télécom, UMR 9189 CRIStAL, Lille, F-59000, France}
\address[painleve]{Univ. Lille, CNRS, UMR 8524 Laboratoire Paul Painlev\'e, Lille, F-59000, France}

% {\tt\small emery.pierson@univ-lille.fr}
%\author[1]{Anonymous submission}% \snm{Surname}\corref{cor1}}
%\cortext[cor1]{Corresponding author: 
%  Tel.: +0-000-000-0000;  
%  fax: +0-000-000-0000;}
%M\emailauthor{example@email.com}{Corresponding Author Name}
%\ead{example@email.com}
    
%\author[2]{Second Author Given Name \snm{Surname}\fnref{fn1}}
%\fntext[fn1]{Footnote 1.}  

%\address[1]{Address, City, Postcode, Country}
%\address[2]{Address, City, Postcode, Country}

%\received{1 February 2017}
\received{\today}
%%%% Do not use the below for submitted manuscripts
%\finalform{28 March 2017}
%\accepted{2 April 2017}
%\availableonline{15 May 2017}
%\communicated{S. Sarkar}

%%%%% This was the first par of the abstract (JC 19/03)
%Comparing 3D surfaces is a challenging problem lying at the heart of many primary research areas in computer graphics, computer vision applications and medical applications. The main difficulty when comparing two triangulated surfaces is that their triangulations do not necessarily have the same number of triangles and, even if they did, there is no natural way to discern what the corresponding triangles would be in each triangulation. The goal of analyzing shapes of surfaces modulo re-triangulations or reparametrizations---their continuous analogues---leads to enormous computational challenges. These are further complicated by the need in many applications to identify surfaces that differ only by Euclidean transformations and similarities.

%%%%%%%%% ABSTRACT
\begin{abstract}
We analyze human poses and motion by introducing three sequences of easily calculated surface descriptors that are invariant under reparametrizations and Euclidean transformations.  These
descriptors are obtained by associating to each finitely-triangulated surface two functions on the unit sphere: for each unit vector $u$ we compute the weighted area of the projection of the surface onto the plane orthogonal to $u$ and the length of its projection onto the line spanned by $u$. The $L_2$ norms and inner products of the projections of these functions onto the space of spherical harmonics of order $k$ provide us with three sequences of Euclidean and reparametrization invariants of the surface. The use of these invariants reduces the comparison of 3D+time surface representations to the comparison of polygonal curves in $\R^n$. The experimental results on the FAUST and  CVSSP3D artificial datasets are promising. Moreover, a slight modification of our method yields good results on the noisy CVSSP3D real dataset. 
\end{abstract}

\begin{keyword}
\KWD 3D Human Shape Analysis \sep 4D Human Retrieval  \sep Convex geometry \sep spherical harmonics analysis
\end{keyword}

\end{frontmatter}

%\linenumbers
%%%%%%%%% BODY TEXT
\section{Introduction}
Comparing 3D surfaces is a challenging problem lying at the heart of many primary research areas in computer graphics, computer vision applications and medical applications. The main difficulty when comparing two triangulated surfaces is that their triangulations do not necessarily have the same number of triangles and, even if they did, there is no natural way to discern what the corresponding triangles would be in each triangulation. The goal of analyzing shapes of surfaces modulo re-triangulations or reparametrizations---their continuous analogues---leads to enormous computational challenges. These are further complicated by the need in many applications to identify surfaces that differ only by Euclidean transformations and similarities. 

A particularly elegant mathematical approach to the problem of comparing surfaces is to 
consider the quotient of the space of embeddings of a fixed surface $S$ into $\R^3$ by the actions of the orientation-preserving diffeomorphisms of $S$ and the group of Euclidean transformations, and provide this quotient with the structure of an infinite-dimensional orbifold. We can then define
and use Riemannian metrics on this orbifold to measure the distance between two given shapes as well as to interpolate between them by computing the (generally unique) geodesic that joins them  ~(\cite{TumpachTPAMI2016}, ~\cite{Kurtekpami2012}). Another exciting approach is that of {\it square root normal fields} or SRNF in which different embeddings and immersions of the surface $S$ modulo translations are described by points in a Hilbert space, and both rotations in $\R^3$ as well as reparametrizations of the surface translate into orthogonal transformations in the Hilbert space ~(\cite{JermynEccv2022}). Both approaches are very general and, in theory at least, permit the perfect or nearly perfect comparison of large classes of shapes. Nevertheless, there are many situations were we would need or prefer a quicker and rougher tool to distinguish, classify, or retrieve shapes from a restricted population of surfaces. An example of such a situation is the subject of this work: the classification and retrieval of human poses and actions. Furthermore, the
articulation of the human body enables it to adopt a great variety of poses with very small changes to the intrinsic geometry of the surface that models it. In flexing an arm or a leg we mostly see small intrinsic changes due to the bulging and stretching of muscles, but the net result in terms of the extrinsic geometry of the body can be substantial. Small changes in the intrinsic geometry may even lead to apparent changes in the genus of the human figure through topological noise when, for instance, hands are clasped or feet and legs are crossed. This points to the unsuitability of approaches that we will call {\it intrinsic,} and which are focused on the metric relations (lengths of curves, angles, and areas) on the surface itself independently of the embedding into the ambient space. 

%% Previous paragraph: added the distinction between intrinsic and extrinsic approaches

\begin{figure}
    \centering
    \includegraphics[width=0.8\linewidth]{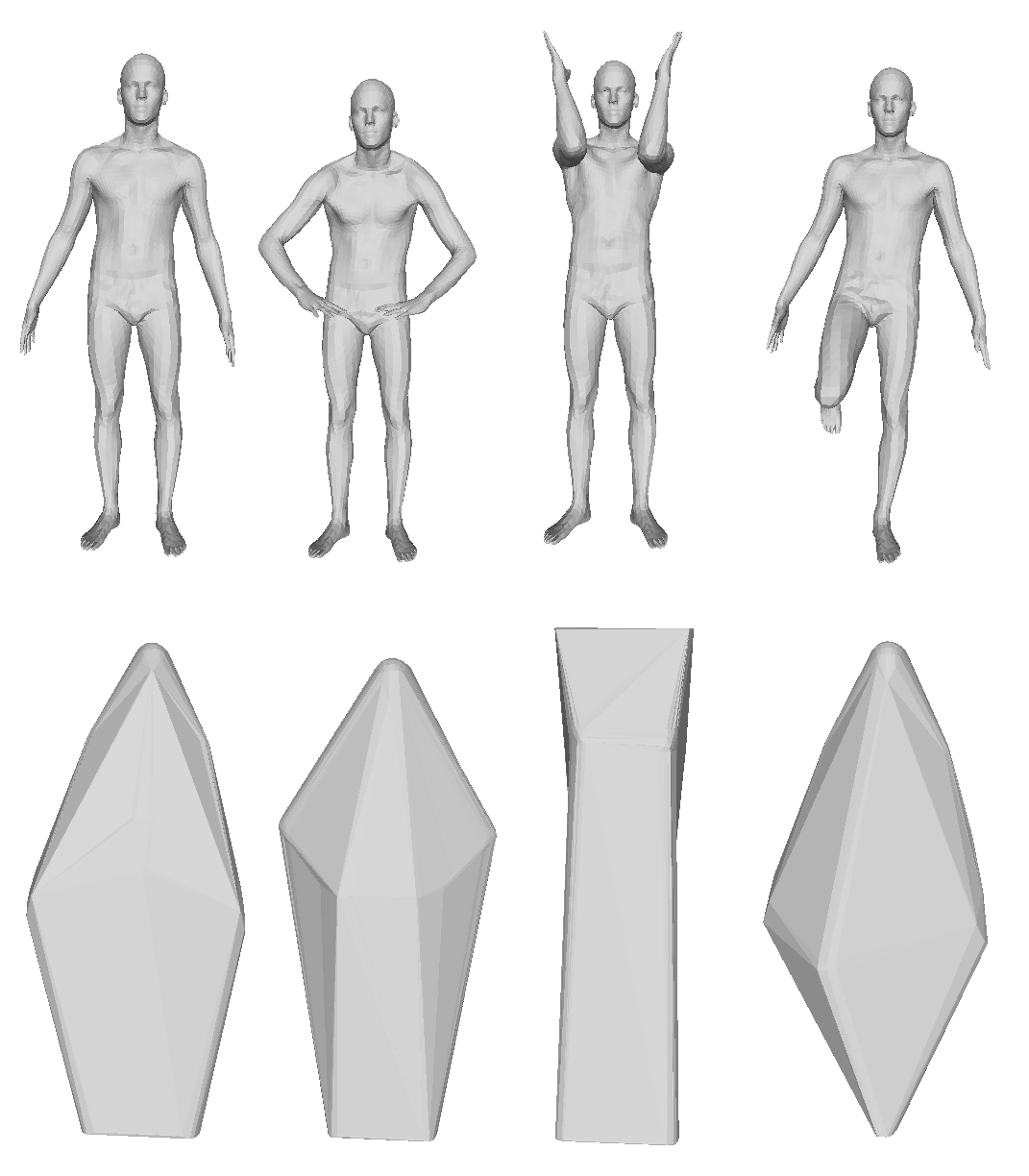}
    \caption{Four human poses from the FAUST dataset along with their corresponding convex hulls.}
    \label{fig:convex_hulls}
\end{figure}

In the analysis and retrieval of human actions we must work with sequences of a hundred human poses, and each pose is represented by a triangulated surface containing thousands or tens of thousands of vertices. This computational complexity is nevertheless offset by the fact that human poses are modeled by a rather restricted population of surfaces. Examination of the databases led us to formulate the hypothesis that a human pose is nearly characterized by its convex hull. The intuition is that if you enclose someone in a tight, perfectly elastic sheet, the different poses of this person will still be distinguishable, or mostly so (see Figure~\ref{fig:convex_hulls}). In 
considering human body motion, where there is a sequence of poses, the probability of recognition of the action from the associated sequence of convex hulls should be even greater, or so the intuition goes. 

This convexity hypothesis led to the idea of considering two of the most basic notions in convex geometry, the convex hull and the surface area measure or extended Gaussian image (EGI), and molding them into three sequences of numerical surface descriptors that are invariant under Euclidean transformations. We do this by first encoding the information of the convex hull in the breadth function, which measures the length of the projection of the surface onto each line passing through the origin, and encoding the information of the EGI in the weighted area function, which for each direction measures the weighted area of the projection of the surface onto the plane perpendicular to it (see Section~\ref{sec:projection} for details). These functions only depend on symmetrizations of the convex hull and EGI (Proposition~\ref{breadth} and Theorem~\ref{weighted-areas}), but are supplementary (i.e., two non-convex surfaces with the same symmetrized convex hull are not likely to have the same symmetrized EGI) and lend themselves nicely to Fourier analysis. Our three sequences of numerical shape descriptors are obtained as the $L_2$ norms and inner products of the projections of these functions onto the space of spherical harmonics of order $k$. In geophysics terminology, these are the power spectra and the cross power spectrum of our two functions (~\cite{Kaula:1967}, ~\cite{Lowes:1974}, and 
see~\cite{KazhdanEG2003} for the introduction of this idea in the context of shape matching). 
%Table~\ref{tab:faust_results} in Section~4 shows that the performance of these descriptors is competitive on the FAUST dataset of~\cite{Bogo:CVPR:2014}. 

%%%%%%%%%%% Next paragraph: analysis vs retrieval of human motion

The main concern of this paper is the problem of analyzing human motion and our numerical descriptors conveniently allow us to reformulate it as a problem of comparing polygonal curves in $\R^n$. In this familiar setting we make use of dynamic time warping (DTW) to compare the curves obtained from the CVSSP3D real and synthetic datasets~\cite{cvssp3d}.

In this paper we did not pay close attention to the effect that noisy data could have on our methods and to the interesting problem of how to make them more robust, but we did test them against the relatively noisy CVSSP3D real dataset (see Table~\ref{tab:cvssp_real_NN}) and remarked that a slight modification to our breadth function to make it more robust yielded good results.

Overall, the contributions of this paper can be summarized as follows.
\begin{itemize}
\item  We present a novel set of descriptors invariant under parametrization, Euclidean transformations, and scaling.
\item We formulate the problem of comparing  sequences of 3D human surfaces as a problem of comparing curves in $\R^n$. Dynamic Time Warping (DTW) is proposed for temporal alignment of these curves. 
\item The method shows promising results for 3D pose and 3D motion retrieval tasks in several datasets. The results are promising  and validate our hypothesis that the analysis of human action can be in good measure reduced to the analysis of sequences of convex hulls of human poses.The experimental results show that our method can be be implemented in a computationally efficient way due to its simple formulation.
 
\end{itemize}

%%%%%%%%%%%%%%%%%%%%%%%%%%%%%%%%%%%%%%%%%%%%%%%%%%%%%%%%%%%%%%%%%%%%%%%%%%%%%%%%%%%

\noindent {\bf Plan of the paper.} In Section ~\ref{sec:related-work}, we review some recent works that have tackled the same or related problems. In Section~\ref{sec:projection} we present the mathematical foundation of our work and the construction of the three sequences of Euclidean and shape 
invariants. This section culminates with the definition of the feature vectors and polygonal curves with which we analyze surfaces and surface motions. The experimental setup is described
in Section~\ref{sec:experiments}. There we present the evaluation criteria, the datasets, and
the results of static pose analysis on the FAUST dataset before moving on to tackle the dynamic analysis of motion in the  CVSSP3D synthetic and real dataset. Finally, we present the mean computation times for the construction of the different polygonal curves associated to the human motions in the various datasets.  Lastly, conclusions and discussion are reported in Section~\ref{sec:conclusion}.
%%%%%%%%%%%%%% In case we really want to give the subsection numbers for everything %%%%%%   
%(\S~\ref{sec:pose_res})
%~(\S~\ref{sec:results})
%~(\S~\ref{sec:real_results})
%%%%%%%%%%%%%%%%%%%%%%%% 

\section{Related Work}
\label{sec:related-work}
%What are the difficulties in performing standard shape analysis on the 3D human data? The main issue is that there is no canonical coordinate system to represent and study the geometry of a nonlinear surface, such as a human.

%\subsection{3D representations.}  
%\textbf{3D  shape decriptors} 
%\todo{such as shape histogram, spin image or spherical harmonics has been used a 3D shape descriptors for shape similarity. It has shown relative success in numerous applications, such as 3D object retrieval. 
%However, when talking about 3D human shapes, the field has been more concentrated on the problem of human shape similarity} . \textbf{Riemannian frameworks}  for analyzing shapes of surfaces for the purposes of matching, comparing, and deforming have proposed by several authors~\cite{Tumpach2016} ~\cite{kurtek2013landmark, kaltenmark2017general}. Even the metrics proposed are invariant under the action of the reparameterization group, the computational costs of these approaches are high in order to be able to apply these frameworks in realistic practical scenarios. 

 %~\cite{zhou20unsupervised} Unsupervised Shape and Pose Disentanglement for 3D Meshes (Zhou et al.) To DO
 
\subsection{Static geometric descriptors}

The challenge in comparing two shapes is to find the best measure of similarity over the space of all transformations. The need for efficient retrieval makes it impractical to explicitly query against all transformations, and two different approaches have been proposed. In the first approach shapes are placed into a canonical coordinate frame (normalizing for translation, scale and rotation) and two shapes are assumed to be aligned when each is in its own frame. Thus, the best measure of similarity can be found without explicitly trying all possible transformations. The second approach describes 3D models through a geometric invariant descriptor so that all transformations of a model result in the
same descriptor. Some descriptors are shown in Table~\ref{tab:SOAT-SP}, which
describes how these methods address translation, scale and
rotation.
%The table ~\ref{tab:SOAT-SP} summarizes some popular approaches to show the diversity of approaches in the literature. We propose 2 types of approaches, one translation invariant, and whenever needed, another fully euclidean invariant. The 2 approaches should be able to capture extrinsic information of human shapes, ie the human pose. 
Other descriptors are {\it intrinsic:} they are defined by local metric properties on the surface itself and, therefore, have natural translation and rotation invariance. They are
 better suited for shape retrieval than for pose retrieval since the intrinsic geometric differences of the surfaces modeling the human body in different poses are not necessarily significant. Examples of these descriptors are HKS, WKS and ShapeDNA, presented in Table~\ref{tab:SOAT-SP}. We refer the reader to~\cite{pickup_shape_2016} for an extensive review and comparison of such descriptors.
%such as Shape Distribution (D2) ~\cite{OsadaACMTGD02}, Spherical Harmonic Descriptor (SHD)~\cite{KazhdanEG2003}

%%%%%%%%%%%%%%%%%%%%%%%%%%%%%%%%%%%%%%%%%%%%%%%%%%%%%%%%%%%%%%%%%%%%%%%%%

\begin{table}
    \begin{center}
    \begin{tabular}{l|c|c|c}
    \hline
    Representation & Tr & Sc & Rot \\ \hline
    \hline
    Shape Distributions~\cite{Osada2001}~\cite{KazhdanEG2003} & I & N & I \\ \hline
    Extended Gaussian Images~\cite{Horn1984}~\cite{KazhdanEG2003} & I & N & N \\ \hline
    Shape Histograms~\cite{AnkerstKKS99}~\cite{KazhdanEG2003} & N & N & N \\ \hline
    Heat Kernel Signatures ~\cite{SunOG09}~\cite{Bronstein2010} & I & N & I \\ \hline
     Wave Kernel Signature ~\cite{Aubry2011} & I & I & I \\ \hline
        ShapeDNA~\cite{ReuterWP06} & I  &  N & I \\ \hline
        GDVAE~\cite{gdvae_2019} (Deep learning) & I  &  N & I \\ \hline
        Neural3DMM~\cite{zhou20unsupervised} (Deep learning) & N  &  N & N \\ \hline
   %      \textcolor{blue}{LIMP ~\cite{LIMP-ECCV2020} (Deep learning)} & ?  &  ? & ? \\ \hline

    \end{tabular}    
    \end{center}
    \caption{A summary of a number of shape descriptors, showing
whether they are (I)nvariant with respect to translation,
scaling and rotation, or  whether they require (N)ormalization.}
    \label{tab:SOAT-SP}
\end{table}

\subsection{Deep Learning}
Deep learning for 3D human poses attracts more and more attention. These new approaches require the reformulation of several deep learning operations, such as regular convolution and pooling/unpooling to the non-regular mesh. Bronstein \etal~\cite{bronstein2017geometric} give a comprehensive overview of the generalization of CNNs on non-Euclidean manifolds. More recently several deep learning approaches propose to learn a latent representation with disentangled shape and pose components. Zhou \etal~\cite{zhou20unsupervised} propose an auto-encoder model that disentangles shape and pose for 3D meshes in an unsupervised manner. However, the proposed neural network requires mesh correspondence, while our approach does not.
%Zhou \etal~\cite{zhou20unsupervised} also show the usefulness of the disentangled codes for the tasks of shape and pose retrieval.
Aumentado-Armstrong \etal~\cite{gdvae_2019} propose a two-level unsupervised Variational Autoencoder (GDVAE), with a disentangled latent space. They utilize point cloud data to learn a latent representation of 3D human shape and thus require training to encode the shape and the pose. They utilize the fact that isometric deformations preserve the spectrum of the Laplace-Beltrami Operator (LBO). The LBO
is a popular way of capturing intrinsic shape. However, the spectrum is very sensitive to noise as shown in our experiments.
%\textcolor{blue}{LIMP ~\cite{LIMP-ECCV2020} uses the idea that change in pose should preserve pairwise geodesic distances. However, the main limitation of LIMP method, lies in the requirement of labeled pointwise correspondences between the training shapes. While our approach does not require such correspondence.}
% However, it is not clear if these methods are invariant to parametrization group. In addition, the extension of deep learning approaches to 4D still remains an open problem.

\subsection{3D shape sequence retrieval} 
%One of the first work on 3D shape sequence classification was made with the objective of view invariant action recognition~\cite{xmas}, along with the popular Xmas dataset.
Huang \etal~\cite{huang2010shape} extended shape distribution, Spin Image, and spherical harmonics to 3D human motion retrieval. These shape descriptors are not necessarily related 
to the geometry of human body. Slama \etal~\cite{slama_3d_2014} propose a 3D human motion analysis framework for shape similarity and retrieval. The shape descriptor, called
Extremal Human Curve (EHC), is a set of 10 curves which connect the extremal points
of the 3D human surface. The authors of~\cite{slama_3d_2014} propose a geometric approach for comparing the shapes of human surfaces via EHC. They exploit the fact that curves can be parameterized canonically
and thus can be compared naturally. However, the need of the detection of extremal points makes this approach sensitive to the noise and to the low quality of the meshes.  In addition, the comparison between pairs of curves increase the computational cost. Another interesting approach is presented by Luo \etal~\cite{luo_spatio-temporal_2016}, where they compute a spatio-temporal graph of 3D Human motion. However, this approach also suffers from being time consuming, and needs the same parameterization along a dataset to perform well.
%In this paper, the authors use similar idea of ~\cite{huang2010shape} to incorporate human motion in the similarity. 
%Unlike ~\cite{huang2010shape} and ~\cite{slama_3d_2014}, we proposed new 3D descriptors invariant to Euclidean transformations and parametrization, fast to  compute, and take into account the geometry of 3D human shape. In addition, we formulated the motion as a trajectory in $\R^n$. However, this approach is sensible to the low quality of the mesh.
In ~\cite{Veinidis19} six static shape descriptors are extracted from each mesh of the human sequence and DTW is used as similarity measure, before proposing to add other information like centroid position and speed. However, some descriptors used in this approach requires a pose normalization for each mesh per frame using two variations of PCA.

%which kind of attention has been deserved to this domain and which problems and applications have been considered, e.g.: [1][2] (but I think there are several other works to consider, and I suggest a careful analysis of the literature)
%\section{Temporal sequence retrieval:} discuss the most related works to this, if anything has been proposed, also for 2D case [3] (notice that 2D technique can always be applied to the 3D case).\\

%In general, Deep learning approaches focus more on 4D reconstruction than on 4D motion classification ~\cite{Mescheder19}.

%At last, 3D human motion data is still quite difficult to access. The classical CVSSP3D~\cite{cvssp3d} and X-mas dataset~\cite{xmas} were for a long time the only data available. The reconstructions are quite challenging compared to more recent 3D human datasets available. The dynamic FAUST dataset~\cite{bogo_dynamic_2017} (designed for shape registration as the original FAUST), and moreover the AMASS~\cite{amass} dataset (reconstructed shapes from human skeletons), are first steps toward more accessible 3D motion datasets.

%\cite{MullerRC05}

%%%%%%%%%%%%%%%%%%%%%%%%%%%% Math section %%%%%%%%%%%%%%%%%%%%%%%%%%%%%%%%%%%%
\section{Projection-based classification of surfaces}
\label{sec:projection}
%%%%%%%%%%%%%%%%%%%%%%% BREADTH REPRESENTATION %%%%%%%%%%%%%%%%%%%%%%%%%%%%%%%%%
\subsection{The breadth representation}
As we mentioned in the introduction, the guiding idea of this paper is that human poses seem
to be determined to a great extent by their convex hulls (see Figure~\ref{fig:convex_hulls}). In 
order to quantify and test this hypothesis, we consider the support and breadth functions of the triangulated surfaces that model the human form.

\begin{definition}
The \textsl{support function} of a set $S \subset \R^n$ evaluated at the unit vector 
$u \in S^{n-1}$ is the quantity
$$
h(S;u) := \sup_{x \in S} \, u \cdot x.
$$
The \textsl{breadth} of the set $S \subset \R^n$ in the direction given by the unit vector 
$u \in S^{n-1}$ is the quantity
$$
b(S;u) := h(S;u) + h(S,-u) = \sup_{x \in S} \, u \cdot x - \inf_{x \in S} \, u \cdot x \ .
$$
\end{definition}

Geometrically speaking, the breadth of a path-connected set in a direction $u$ is simply the length of the orthogonal projection of the set onto a line parallel to $u$. As the following classic result shows, the support function is a way to encode the convex hull. 

\begin{proposition} \label{breadth}
Two sets $S_1, S_2 \subset \R^n$ have the same support function if and only if their convex hulls are equal. Their breadth functions are the same if and only the convex hulls of the
sets $S_1 - S_1$ and $S_2 - S_2$ are equal.
\end{proposition}

\noindent \textsl{Proof.} The convex hull of a set is the intersection of all half-spaces that
contain it. From the definition of the support function, for each unit vector $u$, the half-space $$
H(S;u) := \{x \in \R^n : u \cdot x \leq h(S;u)\}
$$
contains $S$ and is minimal in the sense that it is the unique half-space that contains $S$ and is contained in $H(S;u)$.  From this perspective, the support function is just a way to encode the set of minimal half-spaces, and thus the set of all half-spaces, that contain $S$. It follows that the support function of a set characterizes its convex hull. 

From the linearity of the functions $x \mapsto u \cdot x$ and the 
definition of support function, we have that if $A$ and $B$ are two subsets of $\R^n$, and $\lambda_1$ and $\lambda_2$ are two positive numbers, then 
$$
h(\lambda_1 A + \lambda_2 B; u) = \lambda_1 h(A;u) + \lambda_2 h(B;u) \textrm{ and }
h(-A, u) = h(A; -u) .
$$
From this we conclude that the breadth function of a set $S$ is also the support function of
$S - S$:
$$
b(S;u) = h(S;u) + h(S; -u) = h(S-S;u) .
$$
Since the convex hull of a set is characterized by its support function, we conclude that the convex hulls of the sets $S_1 - S_1$ are the same if and only if the breadth functions of 
$S_1$ and $S_2$ are equal.
 \qed
\medskip

Unlike the breadth function, the support function is not invariant under translations. This can be fixed by moving the center of mass to the origin. Generally speaking, there is less loss of information when working with the support function than with the breadth function, and this should come up in comparing surfaces that have a central symmetry to
those that do not. However, for comparing human figures this did not seem to be the case and we made the choice to work with the breadth function to keep within a geometric tomography framework of studying human shapes through their projections onto lines and planes.

Using that triangles are convex and that the functions $x \mapsto u \cdot x$ $(u \in S^2)$ are linear, the breadth of a triangulated surface $M \subset \R^3$ can be easily computed from just the knowledge of its vertex points $x_1,\ldots,x_N$:
$$
b(M;u) := \max_{1 \leq i \leq N} \, u \cdot x_i - \min_{1 \leq i \leq N} \, u \cdot x_i \ .
$$

%%%%%%% Areas %%%%%

\subsection{Area representation}
Another classical descriptor of convex bodies and surfaces is the \textsl{surface area measure} or, as is better known in computer vision, the  \textsl{extended Gaussian image} (EGI). This is the push-forward of the two-dimensional Hausdorff measure of the surface onto the unit sphere under the Gauss map. For a triangulated surface, we can give a more pedestrian equivalent formulation:

\begin{definition}
Given an oriented triangulated surface $M \subset \R^3$ formed by a union of triangles $T_1 \ldots, T_m$, its extended Gaussian image is the measure on the unit sphere 
$$
\mu_M := \sum_{i=1}^m \textrm{area}(T_i) \, \delta_{n_i} ,
$$
where $n_i$ is the unit vector perpendicular to $T_i$ in the sense defined by the orientation of the surface and $\delta_{n_i}$ is the delta measure concentrated at $n_i$. 
\end{definition}

There are a number of ways to extract feature vectors from the EGI of a surface. We can, for
instance, manufacture them from the moments or the Fourier transform of this measure, but in this work we chose a more intuitive descriptor: the \textsl{weighted area function.}

\begin{definition}
Given an oriented triangulated surface $M \subset \R^3$ formed by a union of triangles $T_1 \ldots, T_m$, its weighted area function is the function on the unit sphere defined by
$$
\mathcal{A}(M;u) := \sum_{i=1}^m |u \cdot n_i| \, \textrm{area}(T_i) ,
$$
where $n_i$ is a unit vector perpendicular to the triangle $T_i$.
\end{definition}

The quantity $\mathcal{A}(M;u)$ is the weighted area of the projection of $M$ onto the plane
orthogonal to $u$. By \textsl{weighted area} we mean that if $k$ different portions of a surface project onto the same piece of plane, the area of this piece is multiplied by $k$. 

Besides being invariant under reparametrizations and translations, the weighted area function is easy to grasp geometrically and very quickly computed. It's relation to the EGI of the surface follows directly from the 
definitions:
$$
\mathcal{A}(M;u) = \int_{S^2} |u \cdot n| \, d\mu_M .
$$
This expression immediately implies that surfaces with the same EGI are indistinguishable by the weighted areas of their projections. Moreover, because the functions $x \mapsto |u \cdot x|$ $(u \in S^2)$
are even, we only see the {\it even part} of the measure $\mu_M$,
$$
\mu^e_M =  \frac{1}{2} \sum_{i=1}^m \textrm{area}(T_i) \, (\delta_{n_i} + \delta_{-n_i}) .
$$
It follows that if the even parts of the surface area measures of two oriented surfaces
are the same, then their weighted area functions are identical. This is all: by a theorem of Choquet (\cite[p. 53]{Choquet:1969}), finite linear combinations of the functions $x \mapsto |u \cdot x|$ $(u \in S^2)$ are dense in the space of even continuous functions on the sphere, and hence if the integrals of all functions of this form with respect to two even measures are the same, the measures must be the same. We summarize:

\begin{theorem}\label{weighted-areas}
Two oriented triangulated surfaces $M_1, M_2 \subset \R^3$ are indistinguishable by the 
weighted areas of their projections if and only if the even parts of their extended Gaussian
images are the same. 
\end{theorem}

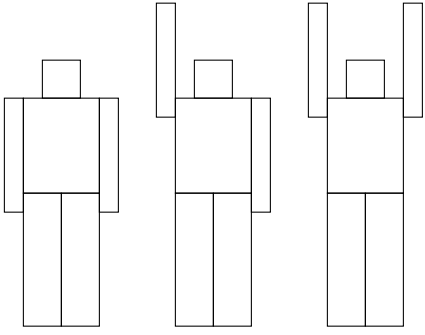
\begin{figure}
    \begin{center}
\begin{tikzpicture}
\draw (0.5,0)rectangle (1,0.5); % head
\draw (0.25,0)rectangle (1.25, -1.25); % body
\draw (0,0) rectangle (0.25, -1.5); % right arm
\draw (1.25,0) rectangle (1.5, -1.5); % left arm
\draw (0.25,-1.25) rectangle (0.75,-3); % right leg
\draw (0.75,-1.25) rectangle (1.25,-3); % left leg

\draw (2.5,0)rectangle (3,0.5); % head
\draw (2.25,0)rectangle (3.25, -1.25); % body
\draw (2,-0.25) rectangle (2.25, 1.25); % right arm
\draw (3.25,0) rectangle (3.5, -1.5); % left arm
\draw (2.25,-1.25) rectangle (2.75,-3);  % right leg
\draw (2.75,-1.25) rectangle (3.25,-3); % left leg

\draw (4.5,0)rectangle (5,0.5); % head
\draw (4.25,0)rectangle (5.25, -1.25); % body
\draw (4,-0.25) rectangle (4.25, 1.25); % right arm
\draw (5.25,-0.25) rectangle (5.5, 1.25); % left arm
\draw (4.25,-1.25) rectangle (4.75,-3); % right leg
\draw (4.75,-1.25) rectangle (5.25,-3); % left leg
\end{tikzpicture}
\caption{Different poses with the same weighted area function, but with different breadth functions.}
\label{fig:width_vs_breath}
%{\bf Fig. 1.} 
\end{center}
\end{figure}

In order to use the weighted area function as a descriptor it is important to understand that if we decompose a surface into a finite or countable number of pieces each of which has a computable area, translating these pieces or flipping them around the origin, and then recomposing them again will give a new surface whose projection onto any plane has the same weighted area as that of the original surface. For instance, if we wish to make use of this technique to classify poses of a human figure  it is useful to keep in mind the following rule of thumb: if we approximate
and decompose the human body as the union of a number of boxes and then these boxes are moved by pure translation and re-glued into a different pose, the method will not effectively distinguish the old and the new poses. An important example is a person standing up with the arms by his/her side and the same person standing up with the arms straight up over his/her head (see Fig.~\ref{fig:width_vs_breath}).

\medskip

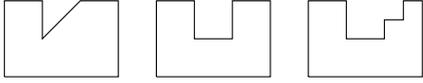
\begin{figure}
\medskip
\begin{center}
\begin{tikzpicture}
\draw (0,0) -- (1.5,0) -- (1.5,1) -- (1,1) -- (0.5,0.5) -- (0.5,1)-- (0,1) -- (0,0);
\draw (2,0) -- (3.5,0) -- (3.5,1) -- (3,1) -- (3,0.5) -- (2.5,0.5)-- (2.5,1) -- (2,1) -- (2,0);
\draw (4,0) -- (5.5,0) -- (5.5,1) -- (5.25,1) -- (5.25,0.75) -- (5,0.75)-- (5,0.5) -- (4.5,0.5) -- (4.5,1) -- (4,1) -- (4,0);
\end{tikzpicture}
\caption{The first two forms have the same convex hull and different weighted area functions, while the last two forms have the same convex hull and EGI.}
\label{fig:width_vs_breath-part2}
\end{center}
\end{figure}

Because of this ``cut-translate-and-paste" invariance, the weighted area may not seem to be as good a descriptor as the breadth, and indeed, that is what our results confirm (see Table~\ref{tab:faust_results}), but it is supplementary information and can be quite discerning in its own right. The weighted area allows us to distinguish some non-convex surfaces that have the same convex hull or breadth function, and although it is possible for two different non-convex surfaces to have the same convex hull and EGI---and, a fortiori, the same breadth and weighted area functions---without being translates
(see Figure~\ref{fig:width_vs_breath-part2} for a simple two-dimensional example of these phenomena), that does not seem to happen to any significant degree in the restricted population of human poses. Nevertheless, the real advantage of considering simultaneously the breadth and weighted area functions will become clearer when we tackle the problem of extracting Euclidean invariants from these functions.

\medskip

%%%%%%%%%%%%%%%%%%%%% EUCLIDEAN AND SHAPE INVARIANTS %%%%%%%%%%%%%%%%%%%%%%%%%%%%%%%%
\subsection{Euclidean and shape invariants}\label{sec:shape-invariants}

In many applications it is not enough to be able to distinguish or classify surfaces up to reparametrizations and translations.  Often we need to do so up to Euclidean transformations or up to similarities. In this section we describe a simple method to extract sequences of Euclidean and shape invariants from the area and breadth function of a surface.

Notice that if $M \subset \R^3$ is a surface and $R$ is a $3 \times 3$ orthogonal matrix, then
$$
\mathcal{A}(RM;u) = \mathcal{A}(M;R^{-1}u) \ \textrm{and} \  b(RM;u) = b(M,R^{-1}u) 
$$
for every unit vector $u$. In other words, the assignments 
$M \mapsto \mathcal{A}(M; \cdot)$ and  $M \mapsto b(M; \cdot)$
are $O(3)$-equivariant maps between the space of surfaces and the space $L_2(S^2)$ of
square-integrable functions on the sphere provided with the usual left $O(3)$-action
$(R,f) \mapsto f \circ R^{-1}$. The classic theory of spherical harmonics (see Lecture~11 of \cite{Arnold:2004}  for a particularly simple description) tells us that this space decomposes into the direct sum
$$
L_2(S^2) = \R \oplus V_1 \oplus V_2 \oplus \cdots ,
$$
where $V_k$ is the $(2k+1)$-dimensional space of spherical harmonics of order $k$ (i.e., 
the restriction to the sphere of homogeneous harmonic polynomials of order $k$ in $\R^3$).
These subspaces are invariant under the action of the orthogonal group and are mutually orthogonal. It follows that if $f$ is a square integrable function on the sphere, we can
decompose $f = f_0 + f_1 + f_2 + ...$ with $f_k \in V_k$, and that the $L_2$ norm of each
component $f_k$, defined by
$$
\|f_k\|_2^2 := \frac{1}{4\pi} \int_{S^2} f_k(u)^2 \,  d\Omega ,
$$
is invariant under the orthogonal group.  Notice that if the function $f$ is an even function, all the odd terms, $f_{2k + 1}$, $k \geq 0$, are zero. This method to extract rotation invariants from spherical functions is classical (see, for instance, \cite{Weyl:invariants}) and is widely used in geophysics (\cite{Kaula:1967, Lowes:1974}), but in the context of computer science it seems to have been introduced in \cite{KazhdanEG2003}, where the term {\it energy representation} of $f$ is used for the sequence $k \mapsto \|f_k\|_2$.

Applying this idea to the area and breadth functions of a surface $M$ we obtain
two sequences of invariants 
$$
 \alpha_k(M) := \ \|\mathcal{A}_{2k}(M; \cdot)\|_2  \ \ \textrm{and} \ \ \beta_k(M) := \|b_{2k}(M; \cdot)\|_2.
$$
To this we add the sequence $\gamma_k(M)$ consisting of the inner products of $\mathcal{A}_{2k}(M; \cdot)$ and $b_{2k}(M; \cdot)$:
$$
\langle \mathcal{A}_{2k}(M; \cdot), b_{2k}(M; \cdot) \rangle_2 =
\frac{1}{4\pi} \int_{S^2} \mathcal{A}_{2k}(M;u) b_{2k}(M;u)   \,  d\Omega ,
$$
which is also a Euclidean invariant of the surface $M$. 

Using the equality
$$
\|f + g \|_2^2 = \|f\|_2^2 + 2\langle f , g \rangle_2 + \|g\|_2^2 ,
$$
we have that 
$$
\gamma_k(M) = \frac{1}{2}\left(\|\mathcal{A}_{2k}(M,\cdot) + b_{2k}(M,\cdot)\|_2^2 - \alpha_k^2(M) - \beta_k^2(M)\right).
$$

It is not clear what is the geometric meaning of most of these invariants, but
by the Cauchy-Crofton formula $\alpha_0(M)$ is simply one-fourth the area of $M$, while $\beta_0(M)$ is $(1/2\pi)$ times the integral of the mean curvature of $M$, {\it provided the surface is convex} (see Chapter~14 in \cite{Santalo:1976}). 

In practice we only know the values of the functions $\mathcal{A}(M; \cdot)$
and $b(M;\cdot)$ on a finite set of grid nodes. Through the use of FFT and cubature
formulas it is possible to numerically compute the invariants $\alpha_k(M)$, $\beta_k(M)$,
and $\gamma_k(M)$ for $0 \leq k \leq l$, where $16(l+1)^2$ is the number of nodes in our grid
(see \cite[pp. 2580--2581]{shtools}). Thus, the $l \times 3$ matrix 
$$
\mathcal{E}_l(M) :=
\begin{pmatrix}
\alpha_0(M) & \beta_0(M) & \gamma_0(M) \\
\vdots & \vdots & \vdots \\
\alpha_l(M) & \beta_l(M) & \gamma_l(M)
\end{pmatrix} ,
$$
which will be our basic Euclidean-invariant representation of the surface $M$, can be effectively computed from the values of the area and breadth functions of $M$ over a uniform sample of $16(l+1)^2$ points on the sphere. 

To end this section we briefly discuss how to extend these Euclidean invariants to 
shape or similarity invariants, where we allow for dilations as well as rotations and translations. To do this we note that if $\lambda$ is a positive real number, then 
$$
\mathcal{A}(\lambda M; u) = \lambda^2\mathcal{A}(M;u) \ \textrm{and} \ 
b(\lambda M; u) = \lambda b(M;u) . 
$$
It follows that 
\begin{align*}
\alpha_k(\lambda M) &= \lambda^2 \alpha_k(M), \ \beta_k(\lambda M) = \lambda \beta_k(M),
 \\
\gamma_k(\lambda M) &= \lambda^3 \gamma_k(M) .    
\end{align*}
We can get rid of the dilation factor in a number of ways. For instance, for each $k \geq 0$,
the quantities
$$
\alpha'_k(M) := \frac{\alpha_k(M)}{||\mathcal{A}(M,; \cdot)||_2} \ \textrm{ and }  \ 
\beta'_k(M) := \frac{\beta_k(M)}{||b(M; \cdot)||_2}  
$$
are shape invariants of $M$, as is
$$
\gamma_k'(M) := \left\|\frac{\mathcal{A}_{2k}(M,\cdot)}{\|\mathcal{A}(M, \cdot)\|_2} + 
\frac{b_{2k}(M,\cdot)}{\|b(M, \cdot)\|_2}\right\|_2 .
$$
As the reader can see, $\gamma_k'(M)$ does not resemble $\gamma_k(M)$ as much as the primed versions of $\alpha_k(M)$ and $\beta_k(M)$ resemble their original versions, but because of the numerical issues we will now discuss, it will be useful for us to have only non-negative shape invariants.

%%%%%%%%%%%%%%%%%%%%%%%%%%%%%%%%%%%%%%%%%%%%%%%%%%%%%%%%%%%%%%%%%%%%%%%%%%%%%%%%%%%%%%%%
\subsection{Numerical considerations}\label{sec:numerical-considerations}
Since the spherical harmonic expansions of the functions $\mathcal{A}(M;\cdot)$ and $b(M;\cdot)$ converge, it follows from Parseval's identity that the invariants $\alpha_k(M)$, $\beta_k(M)$, and $\gamma_k(M)$ tend to zero. They would even decay exponentially if the functions were smooth (see \cite[p.~1151]{Livermore:2012} for a quick proof). In fact, neither function is smooth: the first is a finite convex sum of the non-smooth functions $u \mapsto |u \cdot n_i|$, and the second is support function of a  polytope, namely the convex hull of the differences of all pairs of vertices in the triangulated surface.  However, experimentally (and perhaps due to the great number and small size of the triangles in our triangulated surfaces) the batch of invariants we computed does exhibit exponential decay.
Therefore, the last rows of our basic Euclidean representation
$$
\mathcal{E}_l(M) :=
\begin{pmatrix}
\alpha_0(M) & \beta_0(M) & \gamma_0(M) \\
\vdots & \vdots & \vdots \\
\alpha_l(M) & \beta_l(M) & \gamma_l(M)
\end{pmatrix} ,
$$
will be nearly all zero for even relatively small values of $l$.  We would prefer to deal with
invariants that decay at a slower rate to give some, but not too much, weight to higher harmonics. To be precise, what worked for us was a $t \mapsto 1/t$ decay. To achieve this we change $\alpha_k'(M)$ for
$$
\alpha_k^s(M) := 
\begin{cases}
-\ln(\alpha_k'(M))^{-1} & \textrm{ if } \alpha_k'(M) > 0 , \\
0  & \textrm{ if } \alpha_k'(M) = 0 .
\end{cases}
$$
Similarly, we change $\beta_k'(M)$ for
$$
\beta_k^s(M) := 
\begin{cases}
-\ln(\beta_k'(M))^{-1} & \textrm{ if } \beta_k'(M) > 0 , \\
0  & \textrm{ if } \beta_k'(M) = 0 ,
\end{cases}
$$
and, lastly, we change $\gamma_k'(M)$ for
$$
\gamma_k^s(M) := 
\begin{cases}
-\ln(\gamma_k'(M))^{-1} & \textrm{ if } \gamma_k'(M) > 0 , \\
0  & \textrm{ if } \gamma_k'(M) = 0 .
\end{cases}
$$

From now on we will be working with the modified shape invariant
$$
\mathcal{E}^s_l(M) :=
\begin{pmatrix}
\alpha^s_0(M) & \beta^s_0(M) & \gamma^s_0(M) \\
\vdots & \vdots & \vdots \\
\alpha^s_l(M) & \beta^s_l(M) & \gamma^s_l(M)
\end{pmatrix}.
$$

%%%%%%%%%%%%%%%%%%%%%%%%%%%%%%%%%%%%%%%%%%%%%%%%%%%%%%%%%%%%%%
\subsection{Representation of surfaces and surface evolution}\label{sec:representations}
The final aim of all the preceding mathematics is to represent surfaces as points
and discrete surface motions as polygonal curves in a suitable feature vector space. We consider two types of representation, both of which are independent of the parametrization of the surface: a translation-invariant representation and a shape-invariant representation.

To obtain a translation-invariant representation of a surface $M$ we take a regular sample of $n$ latitude angles, along with a regular sample of $n$ longitude angles of the sphere. We combine them to obtain a spherical grid with $n^2$ nodes $u_1, \ldots, u_{n^2}$ and represent $M$ by one of the following vectors:
\begin{enumerate}
    \item The {\it breadths} feature vector
    $$
    \left(b(M;u_1), \ldots,  b(M;u_{n^2})  \right) \in \R^{n^2}.
    $$
    \item The {\it areas} feature vector
    $$
    \left(\mathcal{A}(M;u_1), \ldots, \mathcal{A}(M;u_{n^2})  \right)  \in \R^{n^2}.
    $$
    \item The {\it areas \& breadths} feature vector which is obtained by joining the
    previous two:
    $$
    \left(\mathcal{A}(M;u_1), \ldots, \mathcal{A}(M;u_{n^2}), b(M;u_1), \ldots,  b(M;u_{n^2})   \right).
    $$
\end{enumerate}
   
To obtain a shape-invariant representation of $M$ we take a similar spherical
grid of $16n^2$ nodes and use the values of $\mathcal{A}(M; \cdot)$  and $b(M;\cdot)$
on these nodes to compute the shape-invariant matrix $\mathcal{E}^s_{n-1}$. Since we wish
to understand how discerning the energies of the breadth and the weighted area
functions are, we shall also consider the first two columns of  $\mathcal{E}^s_{n-1}$ separately. This gives us three shape-invariant feature  vectors:
\begin{enumerate} 
\setcounter{enumi}{3}
\item The {\it area spectrum:} 
$$
(\alpha^s_0(M),\ldots, \alpha_{n-1}^s(M)).
$$
\item The {\it breadth spectrum:} 
$$
(\beta^s_0(M), \ldots, \beta_{n-1}^s(M)).
$$
\item The shape invariant $\mathcal{E}^s_{n-1}$. 
\end{enumerate}

{\it In this paper we will set $n = 8$} and hence when dealing with translation-invariant feature vectors we will be working either in $\R^{64}$ or $\R^{128}$, and when dealing with shape-invariant feature vectors we will be working either in $\R^8$ or $\R^{24}$. In all cases
we will be using the standard Euclidean metric in these spaces to compare surfaces through their
associated vectors. 

In order  to analyze human motion, we need to find
a representation for a sequence of surfaces with timestamps, $(M_0, t_0),\ldots, (M_p,t_p)$.
Using any one of the six feature vectors described above we associate to this sequence a
parametrized polygonal curve in a feature vector space: if $f(M)$ denotes our feature vector,
we construct  the polygonal curve whose vertices are
$\mathbf{x}_j := f(M_j)$,
and for which the parametrization in each segment $\mathbf{x}_j \mathbf{x}_{j+1}$
is given by 
$$
t \longmapsto \frac{t-t_{j+1}}{t_j-t_{j+1}} \, \mathbf{x}_j +  
\frac{t-t_{j}}{t_{j+1}-t_{j}} \, \mathbf{x}_{j+1}  
$$
for $ t_j \leq t \leq t_{j+1}$ and $ 0 \leq j \leq p-1$. 

By this procedure the problem of comparing two human motions, or any other two discrete surface motions, is then reduced to that of choosing a suitable feature vector and comparing the two parametrized polygonal curves associated to the motions. 

\section{Experiments}\label{sec:experiments}

\subsection{Evaluation setup}\label{sub:eval}
%%%%%% To be replaced by diagram?? 
We test the usefulness of the proposed descriptors in two applications: static 3D human pose and 3D human motion retrieval.

%We proceed retrieval experiments on 2 types of data to evaluate our feature vectors chosen between the six defined in\S~\ref{sec:representations}: static 3D human pose, and 3D human motions.
\noindent \textbf{Metric evaluation.} We use three evaluation measures. For all measures a high score implies better results.
\begin{enumerate}
\item \textbf{Nearest neighbor (NN):} It equals one if the nearest neighbor is of the same class of the query, 0 otherwise. This statistic provides an indication of how well a nearest neighbor classifier would perform.
\item \textbf{First-tier (FT), Second-tier (ST):} the percentage of models in the query’s class $C$ that appear within the top $K$ matches, $K$ depending on query’s class size. For a class with $|C|$ members, $K=|C|-1$  for the first tier, and $K=2 \times(|C|-1)$ for the second tier.
\end{enumerate}
The score displayed in evaluation tables are the mean scores computed over the dataset.

%For evaluation of a given representation in retrieval tasks, the usual evaluation scores are as follow: 
%\begin{itemize}
%    \item Nearest Neighbour (NN): The percentage of queries where the closest match belongs to the query class.
%    \item First Tier (FT) : The recall for the (C-1) closest matches were C is the cardinality of the query’s class.
 %   \item Second Tier (ST) : The recall for the 2*(C-1) closest matches were C is the cardinality of the query’s class.
%\end{itemize}
%\subsubsection{Similarity in temporal sequences}
%A framework for analyzing the performance of static descriptors for temporal sequence is presented and used for motion analysis in~\cite{huang_study_2007} and~\cite{huang2010shape}. 
%\paragraph{\textcolor{red}{Temporal ground truth}} When the models are artificial, one can have between shape similarity ground truth, since we have correspondences between vertices. We can construct for each sequence a similarity matrix $R_{TGT}(i, j)$ which contains 1 when $i^{th}$ and $j^{th}$ frame are considered similar, 0 otherwise.

%\paragraph{ROC Curve}
%Based on either the static frame by frame descriptor of its temporal filtered version, we can compute the descriptor similarity matrix $R_S(i,j)$, which contains 1 when $i^{th}$ and $j^{th}$ descriptor are considered similar, 0 otherwise.
%For each motion, we then have a ground truth classification of frames along with the classification coming from the descriptor. We can trace the corresponding ROC curve for each motion.

\subsection{Datasets}
%\paragraph*{Statistical shape dataset}
%The widely used statistical shape dataset collected in~\cite{hasler2009statistical} is composed of human shapes gathered from 144 subjects, which are found in several poses. The 3D surfaces are reconstructed from a statistical model presented in~\cite{hasler2009statistical}. Figure~\ref{fig:stat_poses} shows four poses from this dataset.
%\begin{figure}
%    \centering
%    \includegraphics[width=0.85\linewidth]{illustration/personexamples_nonum.png}
%    \caption{Four poses from the statistical shape dataset.}
%    \label{fig:stat_poses}
%\end{figure}
%The shapes are all aligned with respect to rotation and scale.
%We selected the $18$ poses that were performed by at least four models (p0, p1, p2, p3, p4, p5, p6, p7, p8, p9, p10, p11, p12, p13, p16, p28, p29, p32).

\paragraph*{FAUST dataset}
The FAUST dataset~\cite{Bogo:CVPR:2014}, originally designed for mesh registrations, consists of 3D scans of 10 subjects in 30 different poses and is divided into training and testing sets. In the training set the 3D surfaces are registered to the SMPL human body template. We use those registrations, which are available for 10 different poses,  as a dataset for static human pose retrieval. Some samples are shown in Figure~\ref{fig:convex_hulls}.

\par 
%We propose a benchmark on static pose retrieval, thanks to the FAUST registrations. 
%\paragraph*{AMASS Dataset - CMU part.}
%The AMASS Dataset~\cite{amass} focuses on making 3D "clean" human body motion data accessible. It regroups several skeleton motion datasets reconstructed in 3D using the SMPL-H human body parametrization~\cite{smpl_h} method (6890 and 13776 faces). We focus here on the CMU Graphics Lab Motion Capture Database part of AMASS. In order to concentrate on the locomotions part, which contains jump, run and walk. One such motion is presented in Figure~\ref{fig:amass_data}. The  relevant baseline for shape representation for this dataset is the SMPL parametrization vector of the body. The sampling of the sequences is set to 120Hz.

%\begin{figure}[!t]
%    \centering
%    \includegraphics[width=0.5\linewidth]{illustration/amass_jump.png}
   % \includegraphics[width=0.8\linewidth]{illustration/amass_walk.png}
%    \caption{Jump motion from AMASS-CMU dataset}
%    \label{fig:amass_data}
%\end{figure}

\paragraph*{CVSSP3D dataset}
The CVSSP3D dataset~\cite{cvssp3d} is a 3D human motion dataset created for surface animation. It contains two parts: (1) a synthetic dataset, which contains artificial surfaces animated using known motion capture sequences, and (2) a real dataset, which contains reconstruction of human motions from video sequences. 
We summarize them as follows:
\begin{itemize}
    \item \textit{Real dataset}. This dataset contains 8 models performing 12 different motions: walk, run, jump, bend, hand wave (interaction between two models), jump in place, sit and stand up, run and fall, walk and sit, run then jump and walk, handshake (interaction between two models), pull. The number of vertices for each model vary around 35000. The sampling of the sequences is set to 25Hz.
    
\begin{figure}
    \centering
    \includegraphics[width=0.55\linewidth]{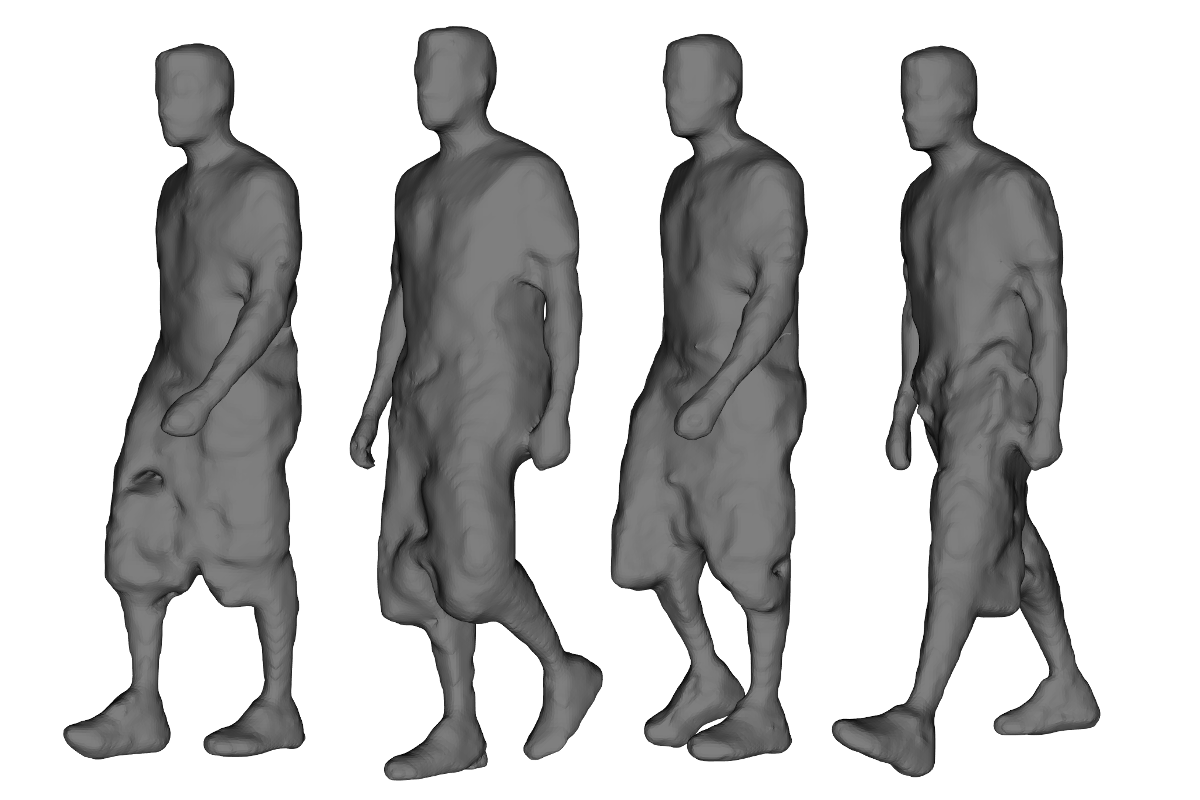}
    \caption{Walking motion from CVSSP3D dataset}
    \label{fig:real_data}
\end{figure}

As the reader can see in Figure~\ref{fig:real_data}, some of the motions in this dataset represent humans moving in loose-fitting clothes. The sensitivity of the reconstructed surface to clothes induces presence of noise in the meshes (see Figure~\ref{fig:noisy}) which makes it a challenge for 3D human motion retrieval.

\item \textit{Synthetic dataset}.
A synthetic model (1290 vertices and 2108 faces) is animated thanks to real motion skeleton data. Fourteen individuals performed each 28 different motions: sneak, walk (slow, fast, turn left/right, circle left/right, cool, cowboy, elderly, tired, macho, march, mickey, sexy,
dainty), run (slow, fast, turn right/left, circle left/right),
sprint, vogue, faint, rock n’roll, shoot. It has already been used~\cite{huang2010shape} for static shape evaluation in the context of 3D motion analysis. A motion from this dataset is presented in Figure~\ref{fig:synth_data}. The sampling of the sequences is set to 25Hz.

\begin{figure}
    \centering
    \includegraphics[width=0.7\linewidth]{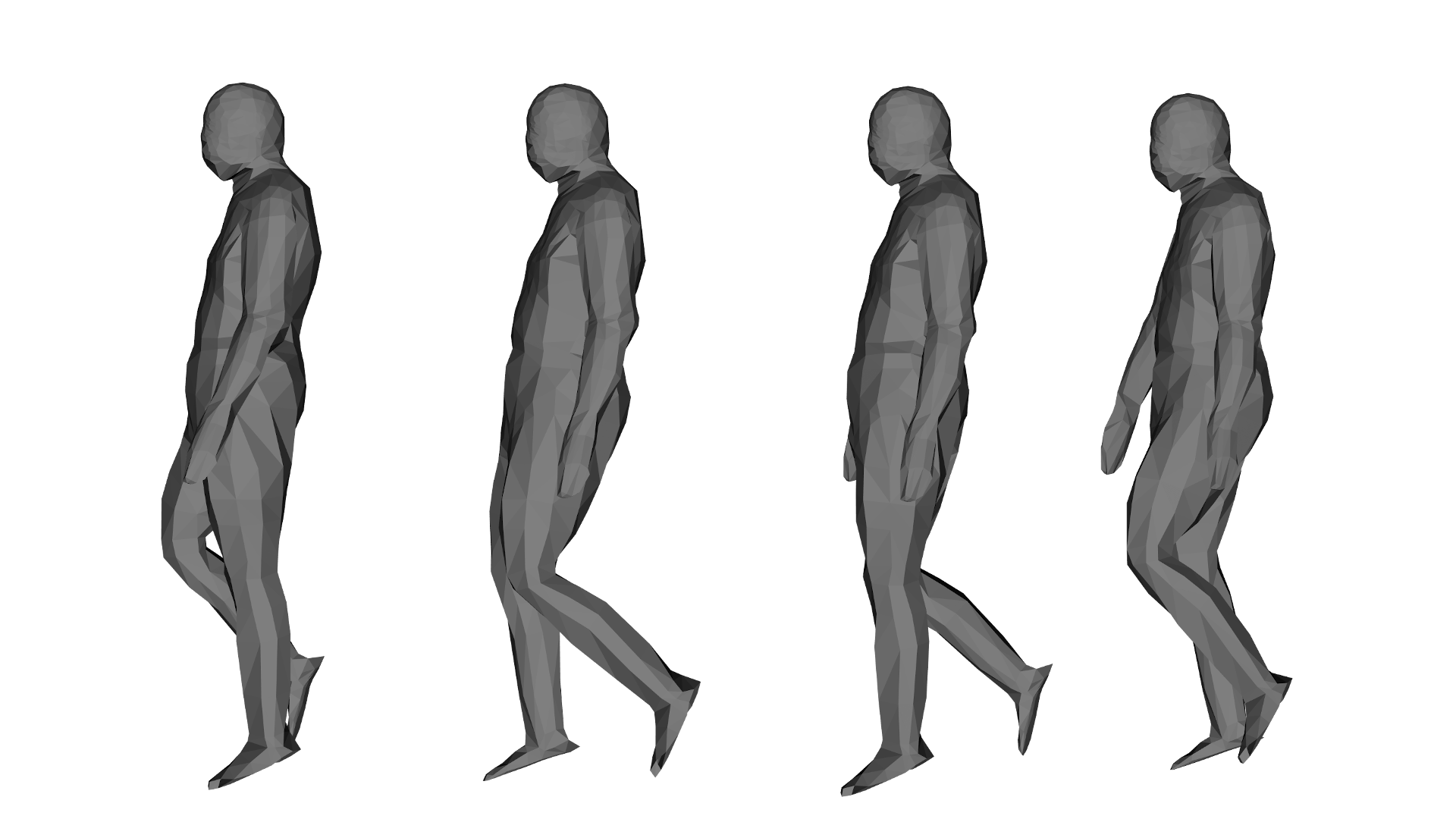}
    \caption{Slow walking motion from CVSSP3D synthetic dataset}
    \label{fig:synth_data}
\end{figure}
\end{itemize}

%%%%%%%%%%%%%%%%%%%%%%%%%%%%%%%%%%%%%%%%%%%%%%%%%%%%%%
\subsection{Static pose retrieval on the FAUST dataset}\label{sec:pose_res}

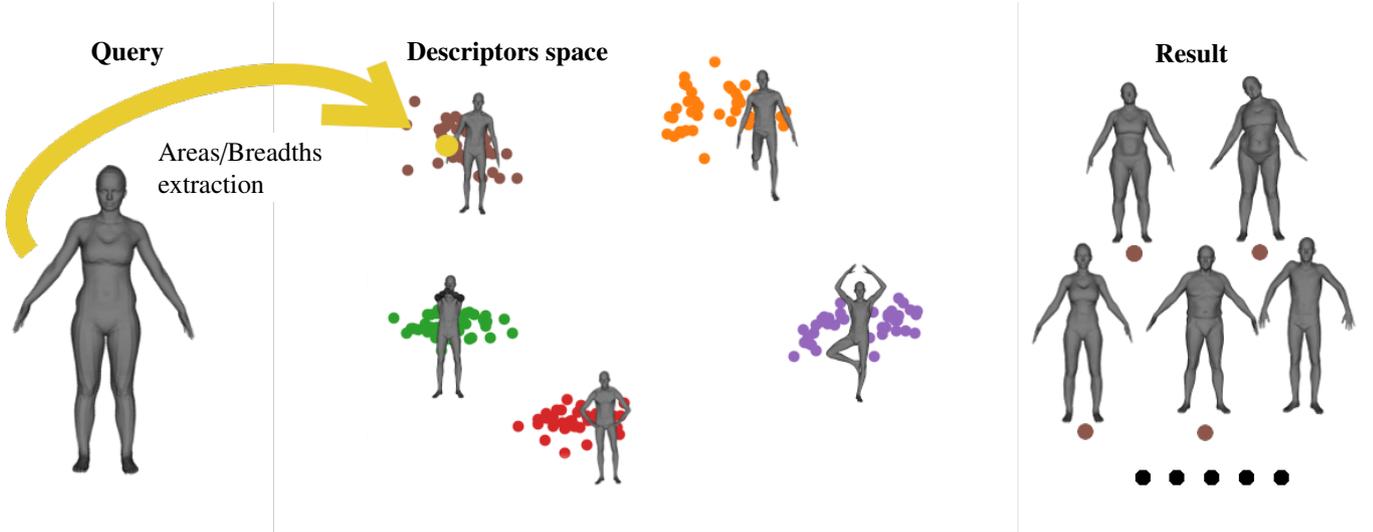
\begin{figure*}[t]

\begin{tikzpicture}
\node (image) at (0,0) {
		\includegraphics[width=\textwidth]{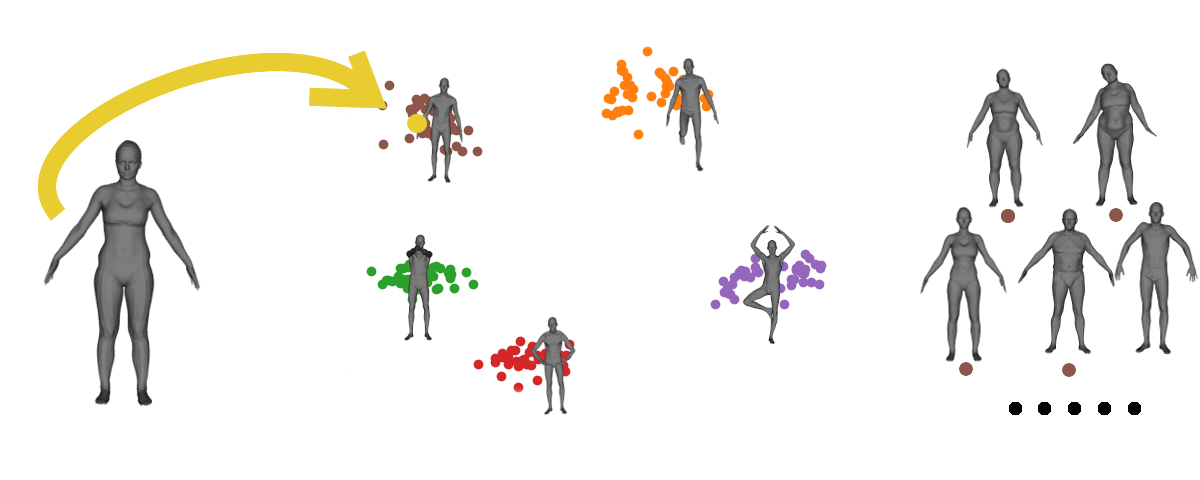}
	};
\node[black,fill=white, text width=2.2cm] at (-5.5, 1.5) {Areas/Breadths \\ extraction};
\node[black,fill=white] at (-2, 3) {\textbf{Descriptors space}};
\node[black,fill=white] at (-7, 3) {\textbf{Query}};
\node[black,fill=white] at (7, 3) {\textbf{Result}};
\end{tikzpicture}
\caption{Overview of our pose retrieval approach: We first compute the descriptors (Areas/Breadths or Areas~\&~Breadths) of all shapes in the database. Given a query shape, we compute its corresponding descriptor and collect the closest shapes in the descriptor space.}
\label{fig:overview_pose_res}
\end{figure*}

Each pose of a dataset is considered as a query belonging to some class. We compute the Euclidean distance between the query pose descriptors and each pose in the dataset (Figure ~\ref{fig:overview_pose_res}).

\noindent \textbf{Comparison with state-of-the-art.} In order to evaluate our descriptor against available methods in the literature, we compare to the following approaches:
\begin{enumerate}
    \item Skinned Multi-Person Linear model (SMPL) pose representation. The SMPL body model~\cite{SMPL:2015} is composed of three parts: a template mesh, a pose vector, and a shape vector. The shape vector represents the (non-rigid) deformation of the template to the shape of the given human body.
    The pose information of a skeletal joint is the relative rotation of the joint of the skeleton compared to its parent joint, and is stored either as the rotation matrix or as axis-angle representation. We convert each joint rotation to quaternion representation as in~\cite{zhou20unsupervised, gdvae_2019} and measure the distance between unit quaternions by $d(q, q') = 1-|q.q'|$. The SMPL body pose vector contains the pose information of 52 joints, and the rotation of the central joint accounts for the global rotation of the shape.
    The representation is a point in $(\mathbb{R}^4)^{51} = \mathbb{R}^{204}$. Due to the construction of the pose vector, this descriptor is rotation invariant. However, this method is time consuming compared to ours because of the needed fitting operation to the mesh.
    \item Aumentado-Armstrong \etal~\cite{gdvae_2019}~propose Geometrically Disentangled VAE (GDVAE), a point cloud variational autoencoder which is trained to disantengle the intrinsic and extrinsic informations of a given shape in the latent space. The authors propose the intrinsic and extrinsic latent vectors for human shape representation.
    % Old sentence and  proposes  a  latent  vector  that  decomposes  in26an  intrinsic  and  an  extrinsic  part.
    We used the FAUST meshes as input of their available trained network, gathered their extrinsic latent vectors (belonging to $\mathbb{R}^{12}$), and used them for human pose retrieval. Although the procedure is parametrization invariant by nature (the networks takes a cloud
    of points as input), the training uses the mesh Laplacian as ground truth information, and this means a constant parameterization along the training set. The network is trained on the SURREAL dataset~\cite{varol17_surreal}  in such a way as to be rotation invariant.
    \item Zhou \etal~\cite{zhou20unsupervised} propose a mesh autoencoder based on the Neural3DMM~\cite{neural3dmm_2019} graph neural network structure. As in the case of GDVAE, this autoencoder disantengles shape and pose in latent space. The network requires that all input meshes have the same parameterization. We apply the FAUST meshes on their available network trained on the AMASS dataset, and use the pose latent vector (belonging to $\mathbb{R}^{112}$) as a descriptor for comparison. Since  the input of the network are the coordinates of the vertices, the approach is not rotation invariant. 
\end{enumerate}

\begin{table}
    \centering
    \begin{tabular}{l|c|c|c}
        Representation & NN & FT & ST \\ \hline \hline 
         Areas & 62 & 50.0 & 67.2 \\\hline
        Breadths & 83 & 63.1 & 76.6 \\\hline
        Areas \& Breadths & \textbf{86} & 67.9 & 80.9 \\\hline \hline
        GDVAE~\cite{gdvae_2019} & 60 & 38.0 & 54.2 \\ \hline
        %GDVAE~\cite{gdvae_2019} pose vector (SMPL) & 55\% & 38.2\%& 54.5\% \\ \hline
        %\cite{zhou20unsupervised} pose vector (SMPL) & 98\% & 88.8\% & 97\% \\ \hline
        Zhou~\etal~\cite{zhou20unsupervised} & 82 & 69.2 & 83.4\\ \hline
        SMPL pose vector & 80 & \textbf{84.4} & \textbf{95.2}\\ \hline
       
   %     Areas (scans) & 62 & 46.0 & 67.2 \\\hline
    %    Breadths (scans) & 85 & 61.8 & 76.5 \\\hline
   %     Areas \& Breadths (scans) & \textbf{90} & 67.3 & 80.9 \\\hline
   %     Areas spectrum* & 52 & 45.8 & 66.0 \\\hline
  %      Breadths spectrum* & 71 & 46.7 & 66.0 \\\hline
   %     Shape invariant* & 72.0 & 54.2 & 72.1 \\\hline
  %      Areas spectrum* (scans) & 34 & 26.8 & 46.0 \\\hline
  %      Breadths spectrum* (scans) & 65 & 53.6 & 60.8 \\\hline
  %      Shape invariant* (scans) & 75 & 51.9 & 69.0 \\\hline
    \end{tabular}
    \caption{Results on pose retrieval for FAUST dataset . 
    %Methods presented with an $\ast$ are rotation invariant methods
    }
    \label{tab:faust_results}
\end{table}

\begin{table}
    \centering
    \begin{tabular}{l|c}
        Representation & Computation time \\ \hline \hline 
          Areas & \textbf{4.1ms}\\\hline
        Breadths & 13.2ms\\\hline
        Areas \& Breadths & 17.2ms\\\hline \hline 
        GDVAE~\cite{gdvae_2019} & 190ms\\ \hline
        Zhou~\etal~\cite{zhou20unsupervised} & 30.7ms\\ \hline
        SMPL pose vector & $\approx 5 min$ \\ \hline 
      
    \end{tabular}
    \caption{Computation time for feature extraction for each method on the FAUST dataset. The computations were performed with NumPy routines on a Intel(R) Core(TM) i5-7600K 3.8GHz CPU, with 8GB of RAM available, except for SMPL, for which the given method needed the use of a GPU.
    %Methods presented with an $\ast$ are rotation invariant methods
    }\label{tab:faust_ctimes}
\end{table}
Table~\ref{tab:faust_results} displays the results obtained for the Areas, Breadths, and 
Areas~\&~Breaths descriptors. The results for the Breadths descriptor is of particular interest as it is here where we see the high correlation between poses and their (symmetrized) convex hull, which validates our main hypothesis. In fact,  Breadth by itself outperforms all previous methods in the NN criterion. When complemented by areas, the performance improves by $3\%$.
The results also show that the SMPL pose vector performs much better for the  FT and ST metrics. This result can be explained by the fact that SMPL has been designed specifically for human shapes. In addition, the SMPL fitting method used here requires a dataset of meshes registered to a template. The Table~\ref{tab:faust_ctimes} shows that our approach is faster than all the methods. It shows also that the computation time of SMPL descriptor is very high.
%we compare the computation time required by our method with that of GDVAE~\cite{gdvae_2019}, Zhou \etal~\cite{zhou20unsupervised} and SMPL pose vector. Our approach is faster than all the methods.
%We compare the previous three representations with our three translation invariant 
%descriptors---Areas, Breadths}, and the combination Areas/Breadths---on the FAUST dataset for static pose retrieval. The results of Table~\ref{tab:faust_results} show that Breadths and Areas \& Breadths are able to outperform learned approaches for pose retrieval for the NN measure, while, by itself,  Areas is less convincing for pose retrieval. 

% The last sentence is not clear and do we really believe what it says we do, or are we just guessing?

%We find our results competitive with the proposed benchmark, our methods yelding better results than~\cite{zhou20unsupervised} in a translation invariant setting (although some of the compared methods are rotation invariant). The actual provided are also consistent with those presented in~\cite{zhou20unsupervised}, where an extensive comparison is yielded with GDVAE, where GDVAE is outperformed by their approach. 

%%% I can't edit this section. I don't understand how the comparison between our methods
%%% and those of the cited paper is being performed and I can't describe how these results
%%% might be reproduced.

\subsection{3D Human Motion retrieval on CVSSP3D artificial dataset}\label{sec:results}
%\subsection{Performance metrics}
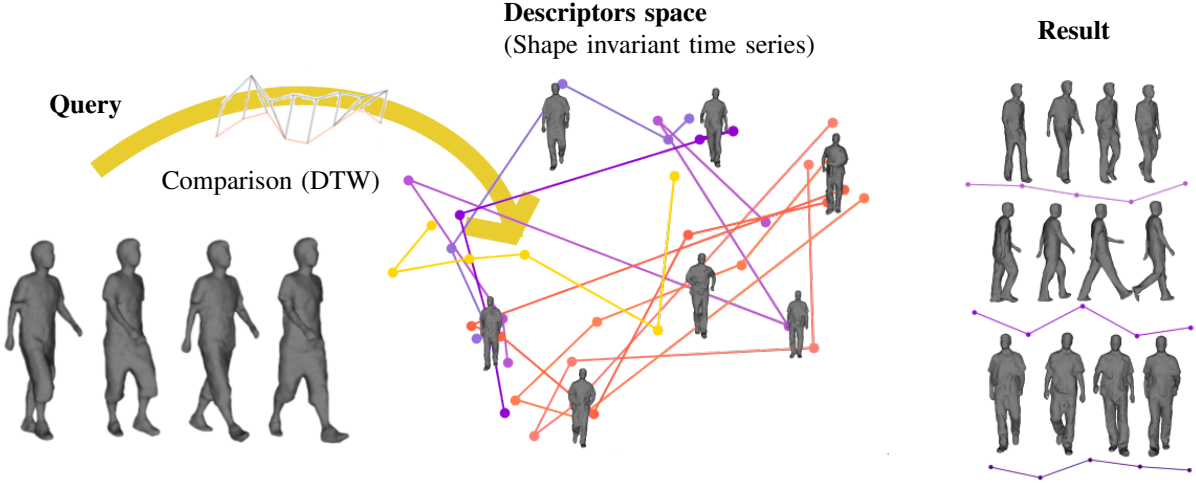
\begin{figure*}[t]
\begin{tikzpicture}
\node (image) at (0,0) {
		\includegraphics[width=0.9\textwidth]{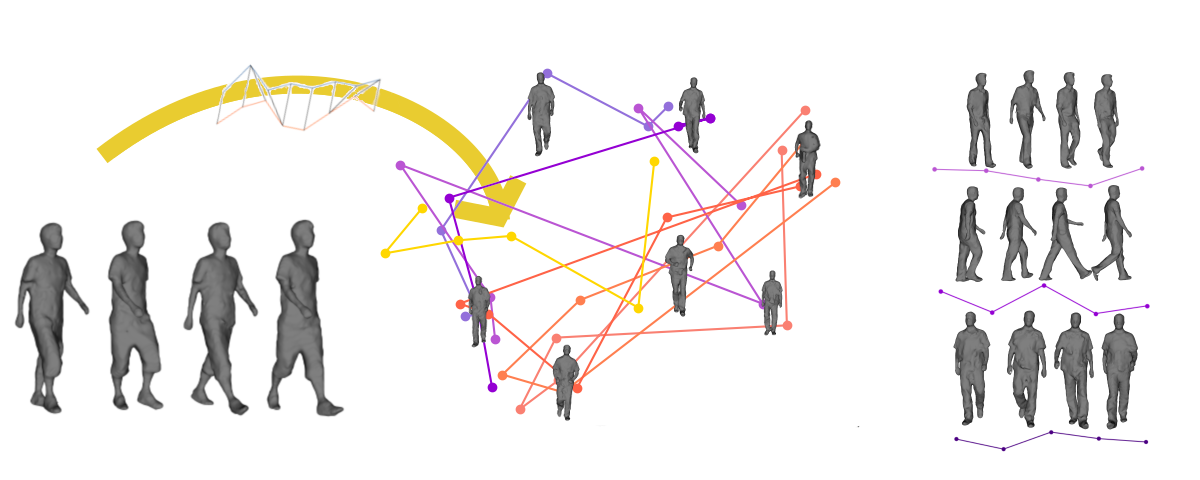}
	};
\node[black,fill=white, text width=3cm] at (-4.5, 1) {Comparison (DTW)};
\node[black,fill=white, text width=5cm] at (1, 3) {\textbf{Descriptors space} \\ (Shape invariant time series)};
\node[black,fill=white] at (-7, 2) {\textbf{Query}};
\node[black,fill=white] at (6, 3) {\textbf{Result}};
\end{tikzpicture}
\caption{Overview of our motion retrieval approach. We first compute the time series of descriptors (areas/breadth spectra or shape invariant) of all motions in the database. Given a query shape, we compute its corresponding time series and compare it against the time series of the database in the descriptor space using dynamic time warping. We then collect the closest motions given this similarity}
\label{fig:overview_motion_res}
\end{figure*}

Each mesh sequence of a dataset is considered as a query belonging to some class. We compute the DTW similarity between the query mesh sequence and each mesh sequence in the dataset (Figure ~\ref{fig:overview_motion_res}).

\textbf{Comparison with state-of-the-art.} An extensive comparison has been made in~\cite{Veinidis19} to evaluate a bench of descriptors for human motion retrieval. The polygonal curves of those descriptors are filtered with a temporal filtering approach(a mean filter is applied along a temporal window of size $K$). Finally, the dynamic time warping distance is used for comparing the resulting curves. 
We compare our invariant descriptors (breadth and area spectrum, shape invariant) to the euclidean and parameterization invariant features presented in~\cite{Veinidis19}, which are:
\begin{enumerate}
    \item Shape Distribution~\cite{OsadaACMTGD02}\cite{Veinidis19} is a 3D descriptor based on pairwise distances. All pairwise distances of a given shape are computed, and the resulting descriptor is an histogram of the obtained distances.
    \item Spin Images~\cite{JohnsonPAMI99}\cite{Veinidis19} is a 3D shape descriptor based on local features. For each point of a shape, a view from the point (the spin image) is computed, which takes the form of a 2D histogram. The resulting descriptor is the sum of all spin images. 
    \item The pretrained GDVAE on SURREAL is applied directly on the dataset. It does not need any supplementary work since the network (PointNet) is parameterization invariant.
    \item The Neural3DMM autoencoder from~\cite{zhou20unsupervised} needs to be specifically trained on the CVSSP3D artificial dataset, since the network is set to specific mesh parameterization and alignment. In order to be fair to the other methods that were not trained on the dataset, we apply a cross identity validation to compute the score. For each individual, we remove its motions from the training dataset. We then compute the retrieval scores for the individual motions using the trained pose representation. The training setting is exactly the same as in~\cite{zhou20unsupervised}.
\end{enumerate}
%\begin{table}
%    \begin{center}
%    \begin{tabular}{l|c|c|c}
%    \hline
%%    Repr. & NN & FT & ST \\ \hline
%    \hline
%    SMPL & \textbf{100} & 88.8 & \textbf{100} \\ \hline
%    Area spectrum & \textbf{100} & \textbf{93.8} & 99.6 \\ \hline
%    Breadth spectrum & \textbf{100} & 89.3 & 99.2 \\ \hline
%    Shape invariant $\mathcal{E}^s_7$ & \textbf{100} & 87.2 & \textbf{100} \\ \hline
%    \end{tabular}    
%    \end{center}
%    \caption{AMASS-CMU dataset results for motion retrieval. We compare our methods against the available SMPL parametrization.}
%    \label{tab:amass_results}
%\end{table}

%The results of our methods on the AMASS-CMU dataset is depicted in  Table~\ref{tab:amass_results}. Note that there is no significant difference between the performance of different features and all of them are competitive compared to the SMPL body pose vector. 
\begin{table}[!h]
    \begin{center}
    \begin{tabular}{l|c|c|c}
    \hline
    Representation & NN & FT & ST \\ \hline
    \hline
     Area spectrum & 81.6 & 56.6 & 68.2 \\ \hline
    Breadth spectrum & \textbf{100} & \textbf{99.8} & \textbf{100} \\ \hline
    Shape invariant $\mathcal{E}^s_7$ & 82.1 & 56.8 & 68.5 \\ \hline \hline
    Shape Distribution~\cite{OsadaACM2002}\cite{Veinidis19} & 92.1 & 88.9 & 97.2 \\ \hline 
    Spin Images~\cite{JohnsonPAMI99}\cite{Veinidis19} & \textbf{100} & 87.1 & 94.1 \\ \hline
    GDVAE~\cite{gdvae_2019} & \textbf{100} & 97.6 & 98.8 \\ \hline
    Zhou \etal~\cite{zhou20unsupervised} & \textbf{100} & 99.6 & 99.6 \\ \hline
    \end{tabular}
    \end{center}
    \caption{CVSSP3D artificial dataset results for motion retrieval using our shape-invariant representations. The results of Shape Distributions and Spin Images are reported from~\cite{Veinidis19}.}
    \label{tab:cvssp_artif_NN}
\end{table}

We report our results on CVSSP3D artificial dataset in Table~\ref{tab:cvssp_artif_NN}.  The window sizes for temporal filtering applied to Shape Distribution and Spin Images are 9 and 8 respectively as in~\cite{Veinidis19}. Our method did not require temporal filtering. We observe that the breadth spectrum has the best performance, near $100\%$, in all criteria.

\subsection{3D Human Motion retrieval on CVSSP3D Real Dataset}\label{sec:real_results}
\begin{figure}
    \centering
    \includegraphics[width=0.9\linewidth]{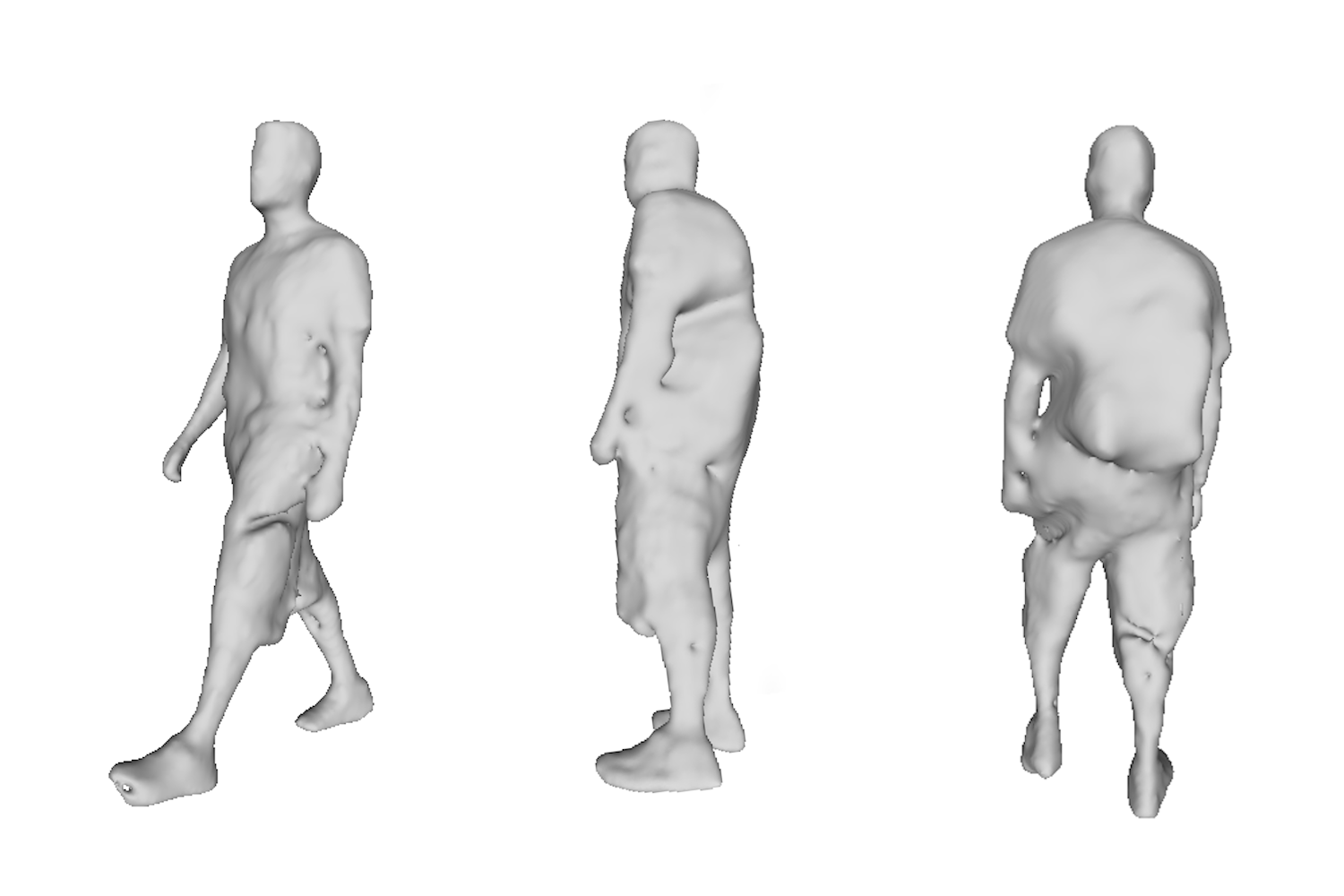}
    \caption{Examples of artifacts in the CVSSP3D real dataset.}
    \label{fig:noisy}
\end{figure}

The CVSSP3D real dataset differs significantly from the artificial human motion dataset because of the relatively noisy data (see Figure~\ref{fig:noisy}) and the various kinds of loose-fitting clothes in some of the models (see Figure~\ref{fig:real_data} and Table~\ref{fig:query}).
This raises the problem of making our descriptors more robust. While a thorough study of this question will be left for a future publication, two conceptually simple and easily implemented modifications to our method can have a significant impact.

\paragraph{The $\lambda$-percentile breadth function.} The breadth function is particularly sensitive to outliers: the maximum  or the minimum value of the function $x \mapsto u \cdot x$  can change significantly with a single noisy vertex $x$. To make this descriptor more robust we make a simple change to the support function of a finite set:

\begin{definition}
Given a finite set $S \subset \R^n$ and a parameter $\lambda$, $0< \lambda \leq 100$, we define the \textsl{$\lambda$-percentile support function} of $S$ as the function  $h_\lambda(S, \cdot)$ that assigns to a unit vector $u \in S^{n-1}$ the $\lambda$-th percentile of the values $\{u \cdot x : x \in S\}$. The \textsl{$\lambda$-percentile breadth function} of $S$ is given by
 $$
 b_\lambda(M;u) = h_\lambda(S;u) + h_\lambda(S;-u). 
 $$
\end{definition}

Defined in terms of the vertices of a triangulation, $b_\lambda(M; \cdot)$ is \textsl{not} invariant under re-triangulations of the surface for $\lambda < 100$. It is only approximately so if the mesh is fine enough and the sizes and shapes of all triangles are comparable. Nevertheless, it is invariant under translations of $M$ and satisfies the equivariance condition
$$
b_\lambda(RM; u) = b_\lambda(M; R^{-1}u).
$$
Provided we understand the conditions on the meshes of the surfaces we are working with, we
can use $b_\lambda(M; \cdot)$ as a  substitute of the breadth function in the construction of shape invariants detailed in Section~\ref{sec:shape-invariants}. We experimented with various values for $\lambda$ and settled on the classic third quartile $\lambda = 75$. We call the
function $b_{75}(M;u)$ {\it Q-breadth.} The analogue of the shape-invariant $\mathcal{E}^s_8$ computed with the Q-breadth function instead of the breadth function will be called the {\it Q-shape invariant.}

\paragraph{Temporal filtering}
Our second trick consists in slightly changing the way we assign polygonal curves to sequences of surfaces with timestamps by making use of a special feature of our invariants.  If we are given a sequence of surfaces we can consider their average breadth function and their average weighted area function, and then proceed with the construction of the feature vectors. Note that for the breadth spectrum, the area spectrum and the shape invariant this is not the same as averaging the feature vectors themselves (we tried that too: the results were not
as good). This particularity of our representation allows us the possibility to perform a simple discrete convolution or temporal filtering on the data: given a sequence of surfaces with timestamps, $(M_0, t_0),\ldots, (M_p,t_p)$ and a number $K$, $0 < K < p$ we consider the timestamped averages of breadth and weighted area functions, which are both represented here by $f$ to avoid redundancy,
$$
\bar{f}_{t_i}(M;u) := \frac{1}{2K + 1}  \sum_{-K \leq j \leq K} f(M_{i + j}; u), \ 
K \leq i \leq p-K .
$$
With the sequence of timestamped averaged functions 
$$\bar{f}_{t_K}(M;u), \ldots, \bar{f}_{t_{p-K}}(M;u)
$$ 
we construct our timestamped feature vectors and the corresponding polygonal curve as described in Section~\ref{sec:representations}. Note that this temporal filtering approach is slightly different from the one proposed in~\cite{Veinidis19} -- our approach is using the specific structure of our descriptors.
The results of our experiments and comparisons on the CVSSP3D real dataset are reported in Table~\ref{tab:cvssp_real_NN}. Again we report the results of Shape distances and Spin Images from~\cite{Veinidis19}. We display in this table the used windows size for temporal filtering of each method. 
For this relatively noisy dataset, the table clearly shows 
the advantage of using the spectrum of the  Q-breadth function and the Q-shape invariant.

The results in Table \ref{tab:cvssp_real_NN} show that the Q-shape invariant outperforms all other methods, including the deep learning method GDVAE whose performance drops significantly in the presence of noise. This can be explained by the noise-sensitivity of the spectrum of the Laplace-Beltrami Operator.

A remarkable difference between the results in Table~\ref{tab:cvssp_real_NN} and those of 
Table~\ref{tab:cvssp_artif_NN} is that the first tier measure is quite low compared to the NN measure for all features. In order to give an idea of how the tier are distributed, a first tier query is illustrated in Table~\ref{fig:query}.

\begin{table}
    \begin{center}
    \begin{tabular}{l|c|c|c|c}
    \hline
    Repr. & $K$ & NN & FT & ST \\ \hline
    \hline
    %Area spectrum & No & 69.5 & 50.7 & 65.5 \\ \hline
    %Area spectrum & 1 & 68.4 & 50.5 & 65.6 \\ \hline
    Area spectrum & 14 & 67.5 & 47.0 & 63.2 \\ \hline
    %Breadth spectrum & No & 54.7 & 33.2 & 46.5 \\ \hline
    %Breadth spectrum & 5 & 58.8 & 33.4 & 46.1 \\ \hline
    Breadth spectrum & 15 & 63.7 & 39.1 & 52.5 \\ \hline
    %Q-breadth spectrum & No & 82.1 & 49.3 & 64.0 \\ \hline
    %Q-breadth spectrum & 3 & 82.1 & 50.3 & 64.8 \\ \hline
    Q-breadth spectrum & 5 & 80.0 & 44.8 & 59.5 \\ \hline
    %Shape invariant $\mathcal{E}^s_7$ & No & 60.0 & 38.6 & 53.7 \\ \hline
    %Shape invariant $\mathcal{E}^s_7$ & Yes & 61.1 & 38.7 & 53.0 \\ \hline
    Shape invariant $\mathcal{E}^s_7$ & 15 & 62.5 & 41.8 & 57.9 \\ \hline
    %Q-shape invariant & No & 83.3 & 57.4 & 73.9 \\ \hline
    %Q-shape invariant & Yes & \textbf{83.3} & \textbf{58.8} & \textbf{73.6} \\ \hline
    Q-shape invariant & 4 & \textbf{82.5} & 51.3 & \textbf{68.8} \\ \hline \hline
    %2D projections & 1 & 66.3 & 56.3 & 75.9 \\ \hline
    %3D spherical & 14 & 92.5 & 72.7 & 86.1 \\ \hline 
    %Hybrid & 6 & 85.0 & 74.5 & 88.9 \\ \hline 
    %PANORAMA \cite{PapadakisIJCV2010} & 15 & 72.5 & 55.0 & 73.0 \\ \hline 
    Shape Distribution~\cite{OsadaACM2002} & 1 & 77.5 & \textbf{51.6} & 65.5 \\ \hline
    Spin Images~\cite{JohnsonPAMI99} & 6 & 66.3 & 43.2 & 59.5 \\ \hline
    GDVAE~\cite{gdvae_2019} & 14 & 38.7 & 31.6 & 51.6\\ \hline
    
    \end{tabular}
    \end{center}
    \caption{CVSSP3D real dataset results for motion retrieval using our shape-invariant representations and their Q-versions. The results of Shape Distributions and Spin Images are reported from~\cite{Veinidis19}. The K value is the best window size for temporal filtering, and the displayed score are the corresponding best scores.}
    \label{tab:cvssp_real_NN}
\end{table}

% \begin{figure*}
%     \centering
%     {\color{blue}
%     \begin{subfigure}{0.2\linewidth}
%     \includegraphics[width=\linewidth]{query/nikos_002.png}
%     \subcaption{Nikos, Walk (Query)}
%     \end{subfigure}
%     \begin{subfigure}{0.2\linewidth}
%     \includegraphics[width=\linewidth]{query/jean_014.png}
%     \subcaption{Jean, Walk}
%     \end{subfigure}
%     \begin{subfigure}{0.2\linewidth}
%     \includegraphics[width=\linewidth]{query/jon_041.png}
%     \subcaption{Jon, Walk}
%     \end{subfigure}
%     \begin{subfigure}{0.2\linewidth}
%     \includegraphics[width=\linewidth]{query/hansung_029.png}
%     \subcaption{Hansung, Walk}
%     \end{subfigure}
%     \begin{subfigure}{0.2\linewidth}
%     \includegraphics[width=\linewidth]{query/chris_028.png}
%     \subcaption{Chris, Walk}
%     \end{subfigure}
%     \begin{subfigure}{0.2\linewidth}
%     \includegraphics[width=\linewidth]{query/haidi_001.png}
%     \subcaption{Haidi, Walk}
%     \end{subfigure}
%     \begin{subfigure}{0.2\linewidth}
%     \includegraphics[width=\linewidth]{query/Hansun_050.png}
%     \subcaption{Hansung, Walk, Run and Jump}
%     \end{subfigure}
%     \begin{subfigure}{0.2\linewidth}
%     \includegraphics[width=\linewidth]{query/nikos_003.png}
%     \subcaption{Nikos, Run}
%     \end{subfigure}
%     \caption{First seven results of a query on the CVSSP3D real dataset using the Q-shape invariant as our representation.}
%     \label{fig:my_label}}
% \end{figure*}

\begin{table}[h!]
  \centering
  \begin{tabular}{ | c | c |}
    \hline
    Motion & Picture \\ \hline
    Nikos, Walk (Query) &
    \begin{minipage}{.2\textwidth}
      \includegraphics[width=\linewidth]{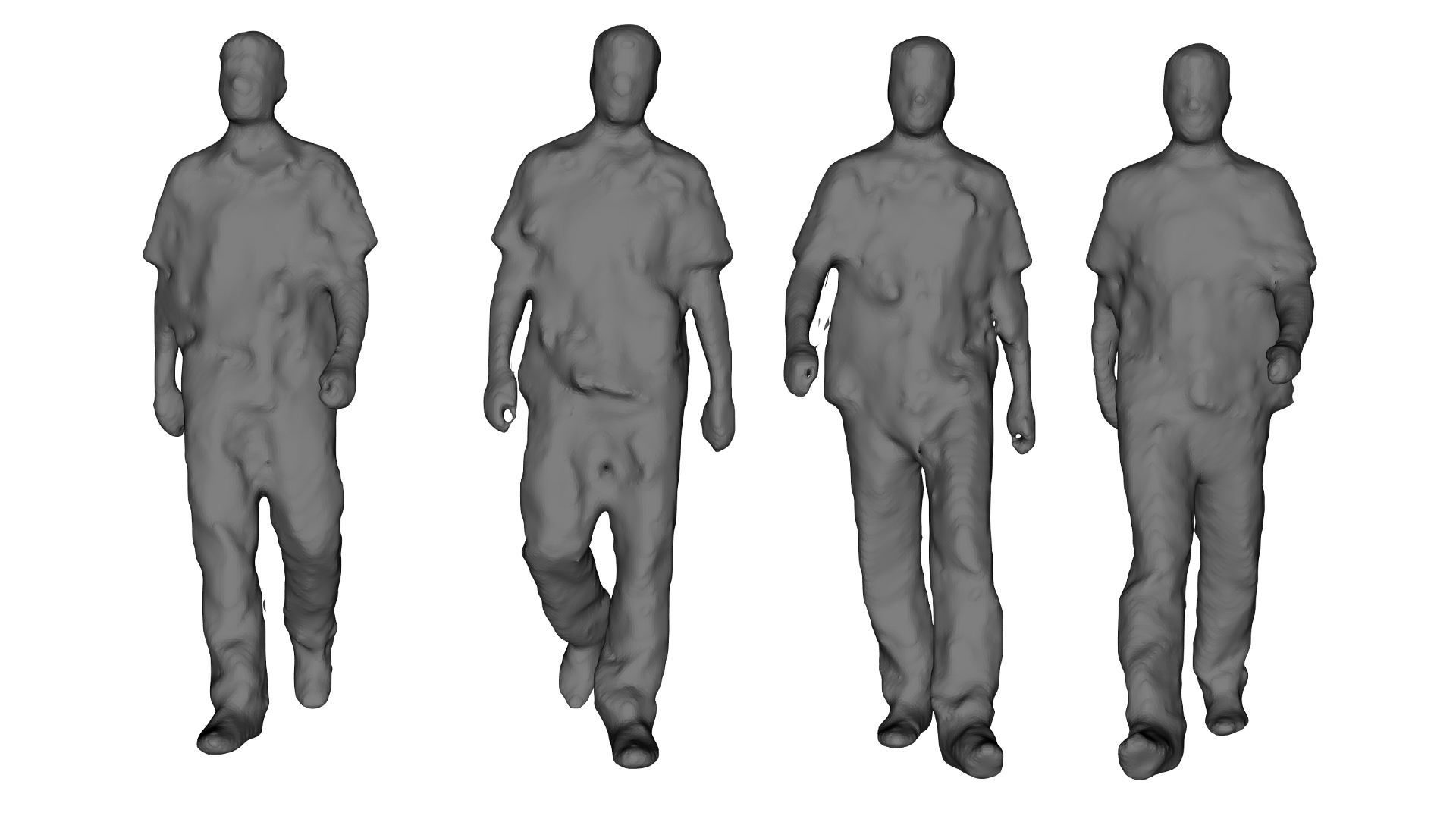}
    \end{minipage}
    
    \\ \hline 
    \hline 
    Jean, Walk&
    \begin{minipage}{.2\textwidth}
      \includegraphics[width=\linewidth]{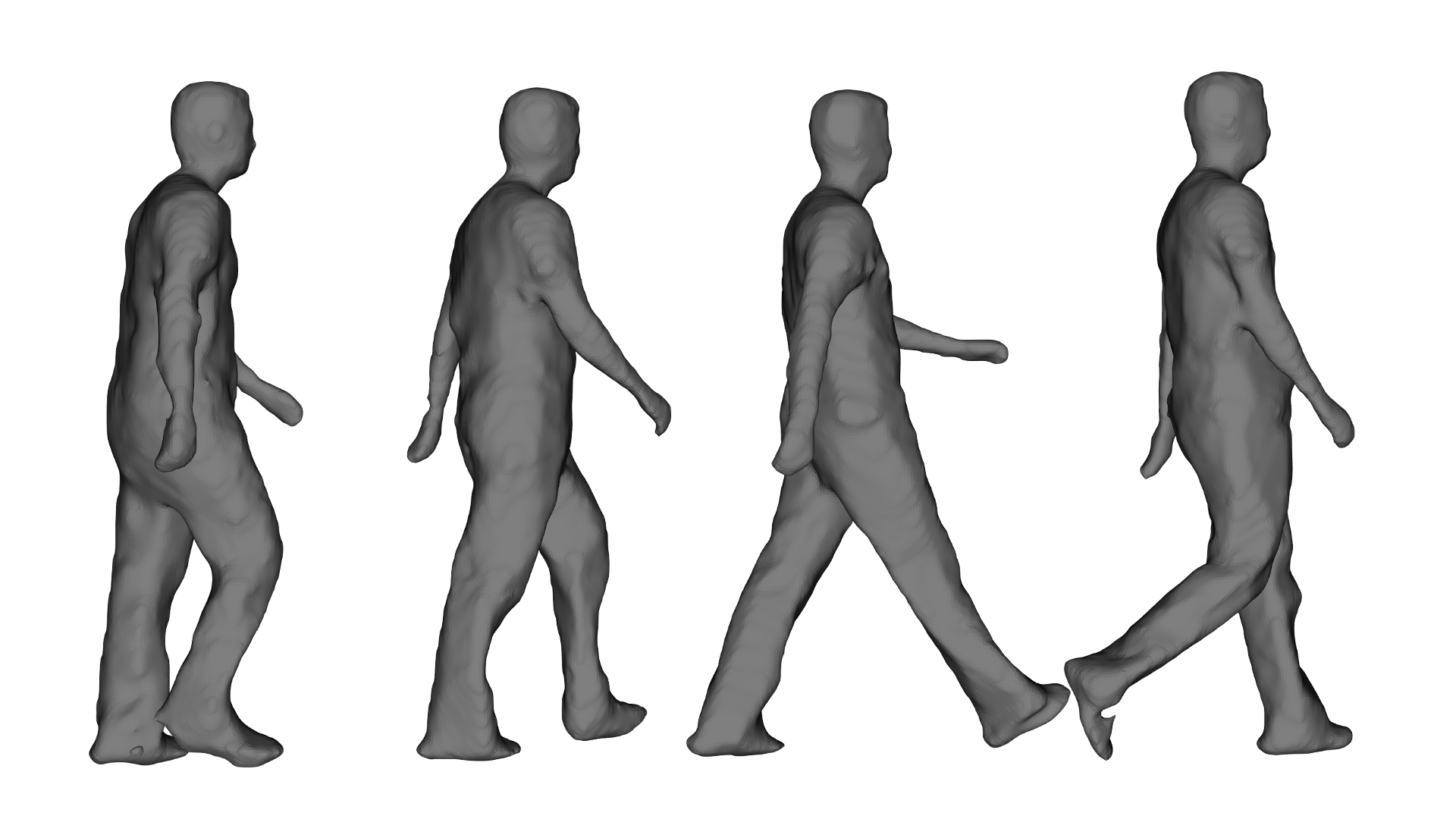}
    \end{minipage}
    
    \\ \hline 
    Jon, Walk&
    \begin{minipage}{.2\textwidth}
      \includegraphics[width=\linewidth]{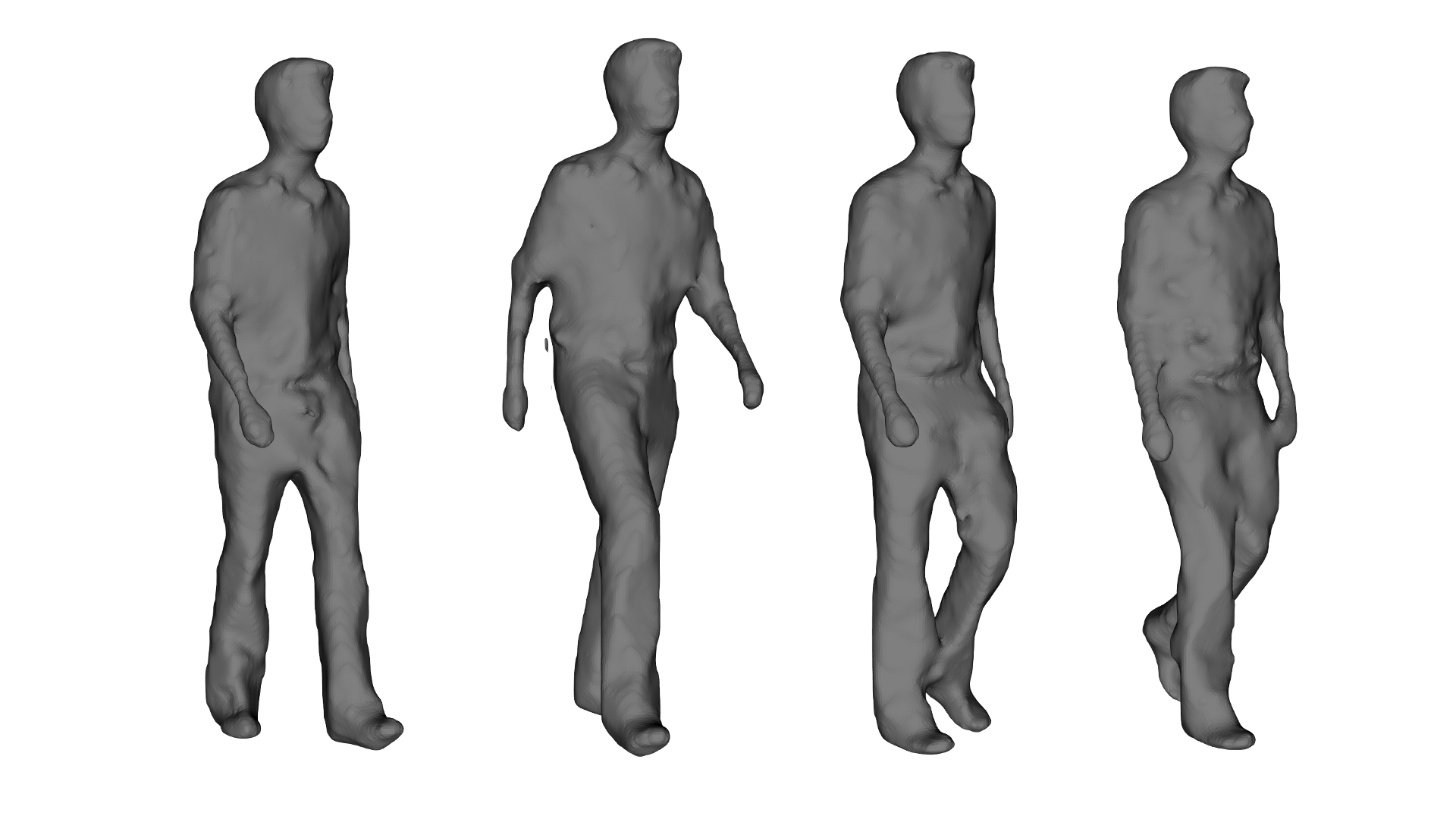}
    \end{minipage}
    
    \\ \hline 
    Hansung
    Walk&
    \begin{minipage}{.2\textwidth}
      \includegraphics[width=\linewidth]{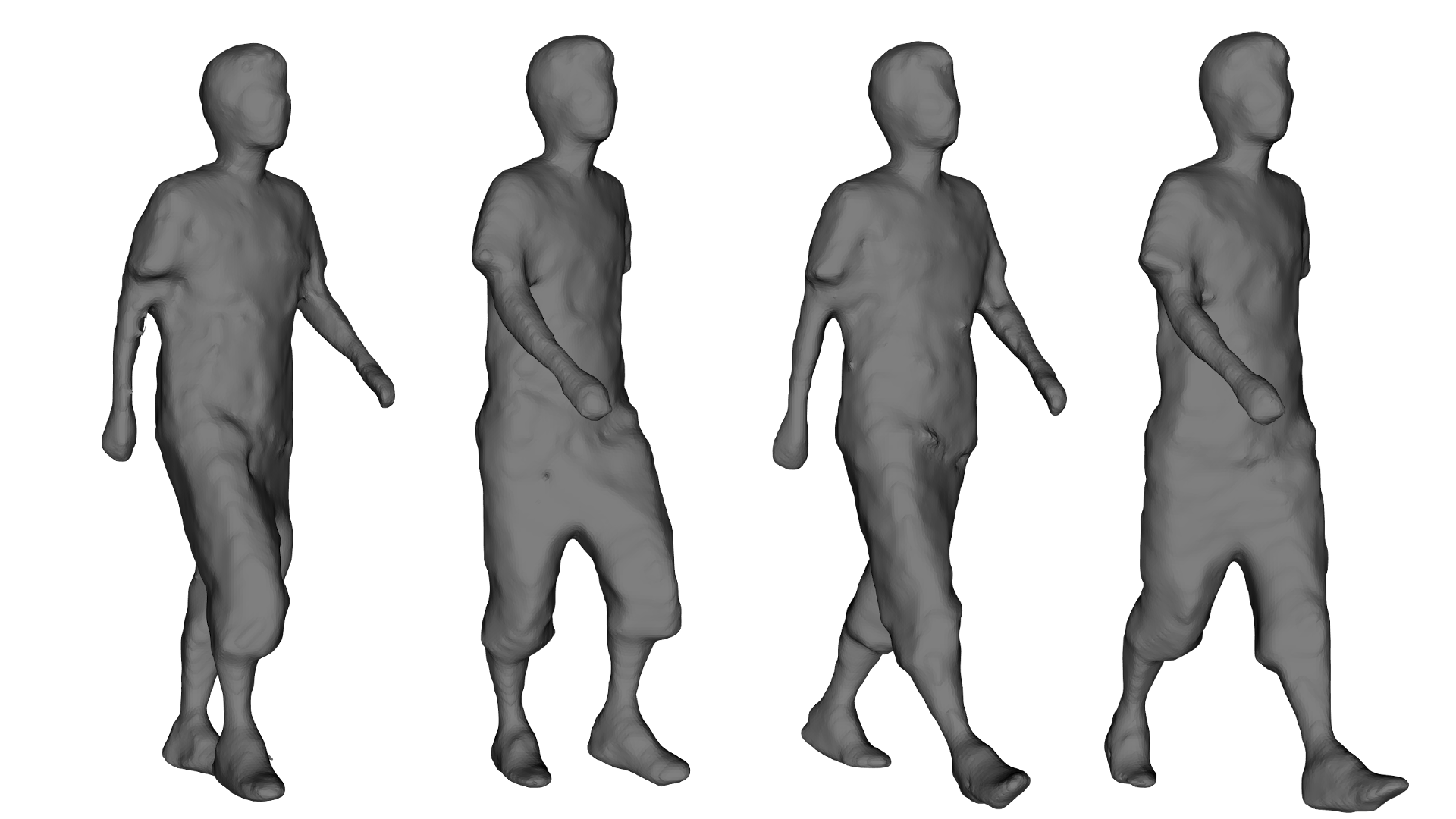}
    \end{minipage}
    
    \\ \hline 
    Chris, Walk &
    \begin{minipage}{.2\textwidth}
      \includegraphics[width=\linewidth]{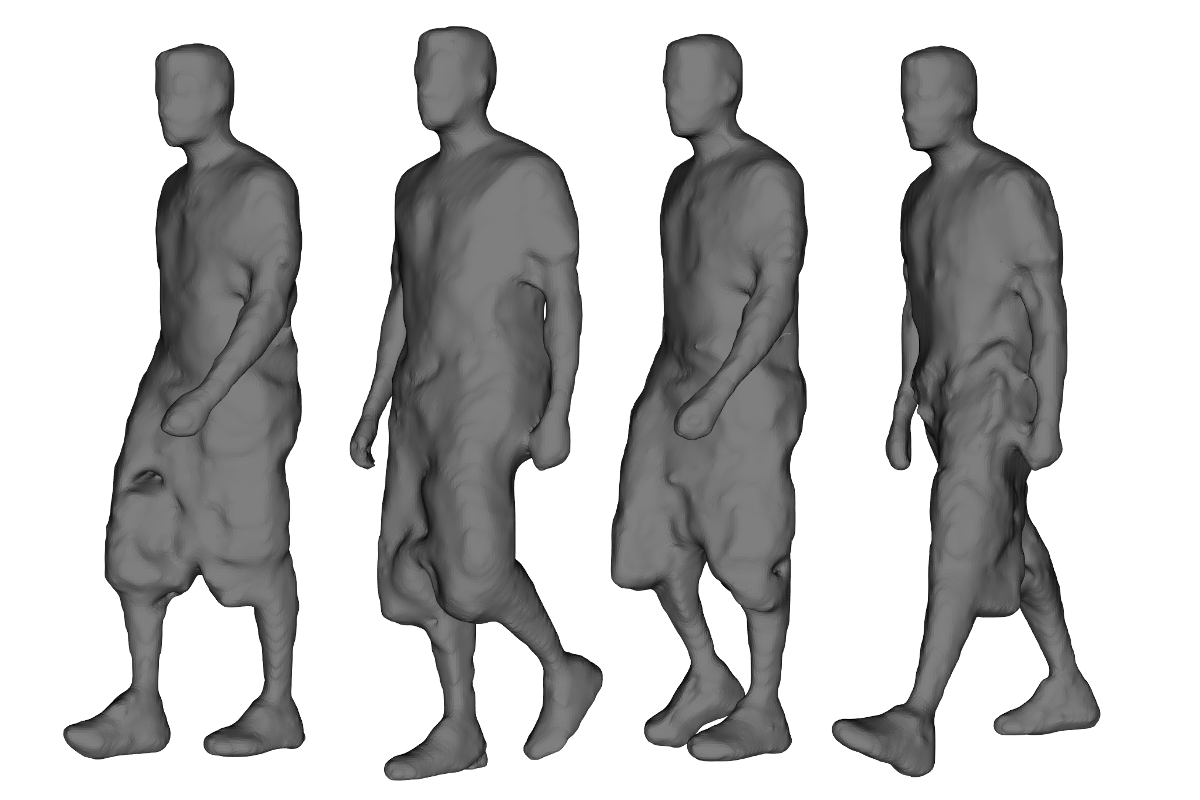}
    \end{minipage}
    
    \\ \hline 
    Haidi, Walk&
    \begin{minipage}{.2\textwidth}
      \includegraphics[width=\linewidth]{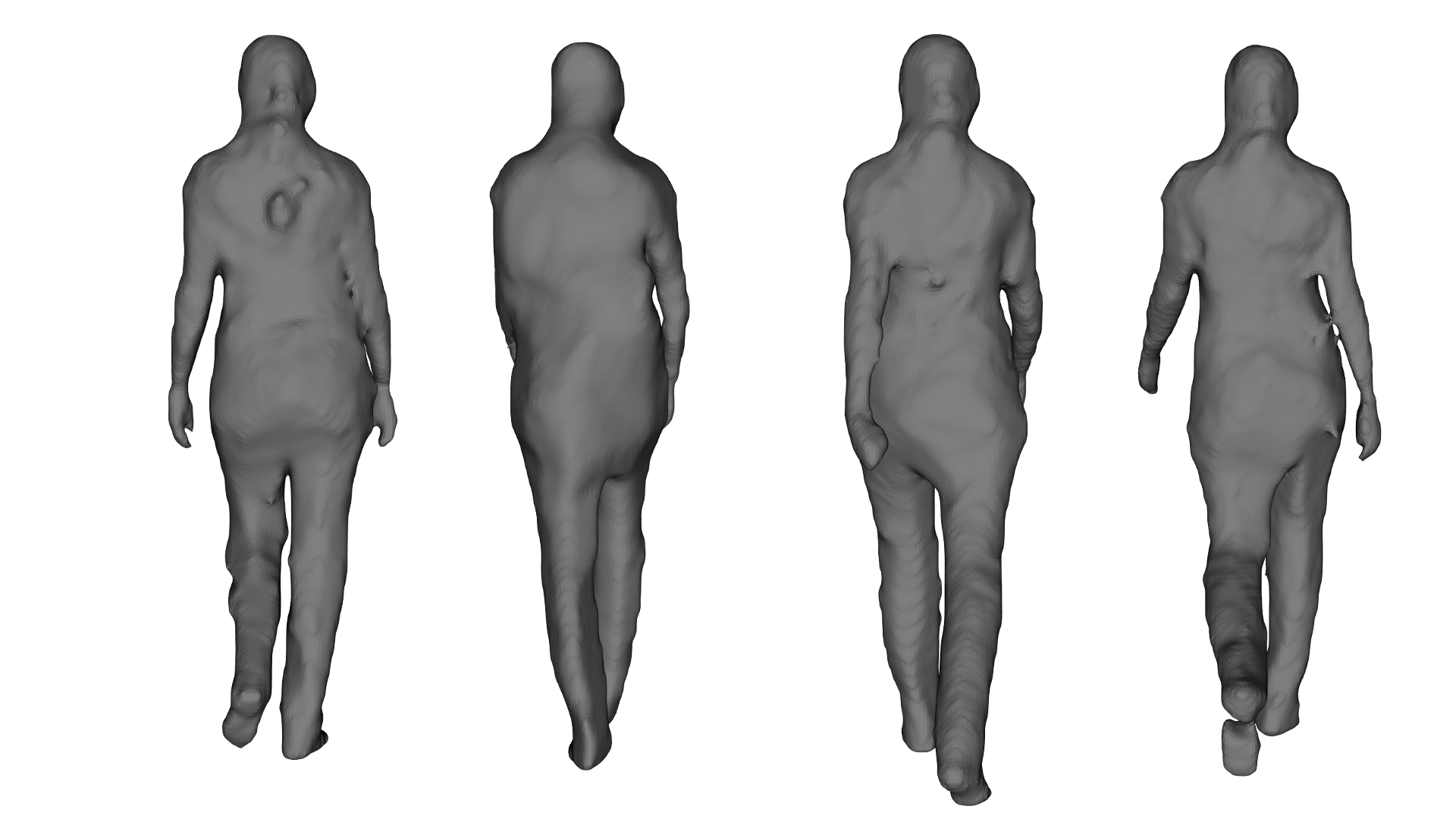}
    \end{minipage}
    
    \\ \hline 
    Hansung, Walk, Run and Jump &
    \begin{minipage}{.2\textwidth}
      \includegraphics[width=\linewidth]{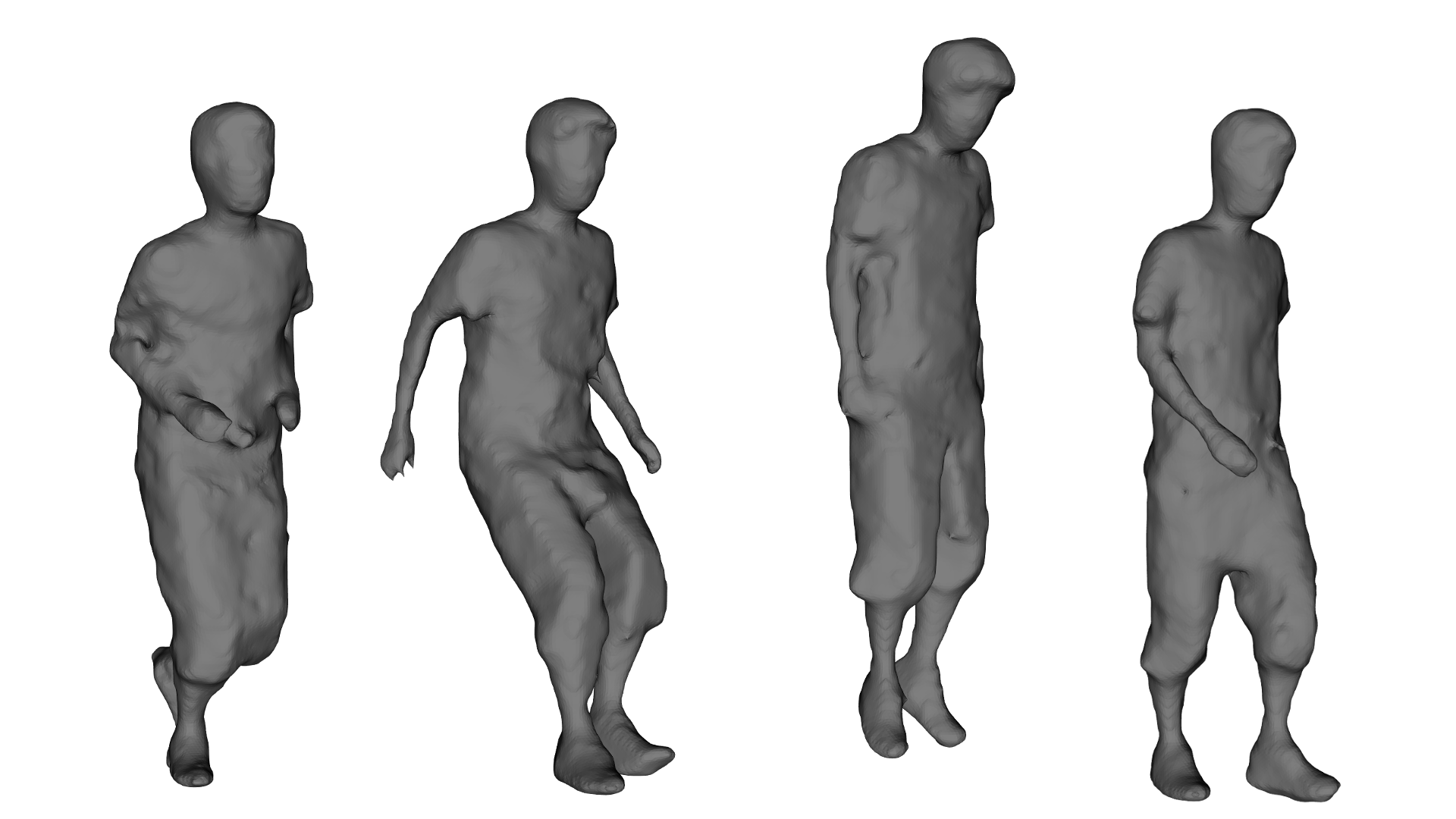}
    \end{minipage}
    
    \\ \hline 
    Nikos, Run&
    \begin{minipage}{.2\textwidth}
      \includegraphics[width=\linewidth]{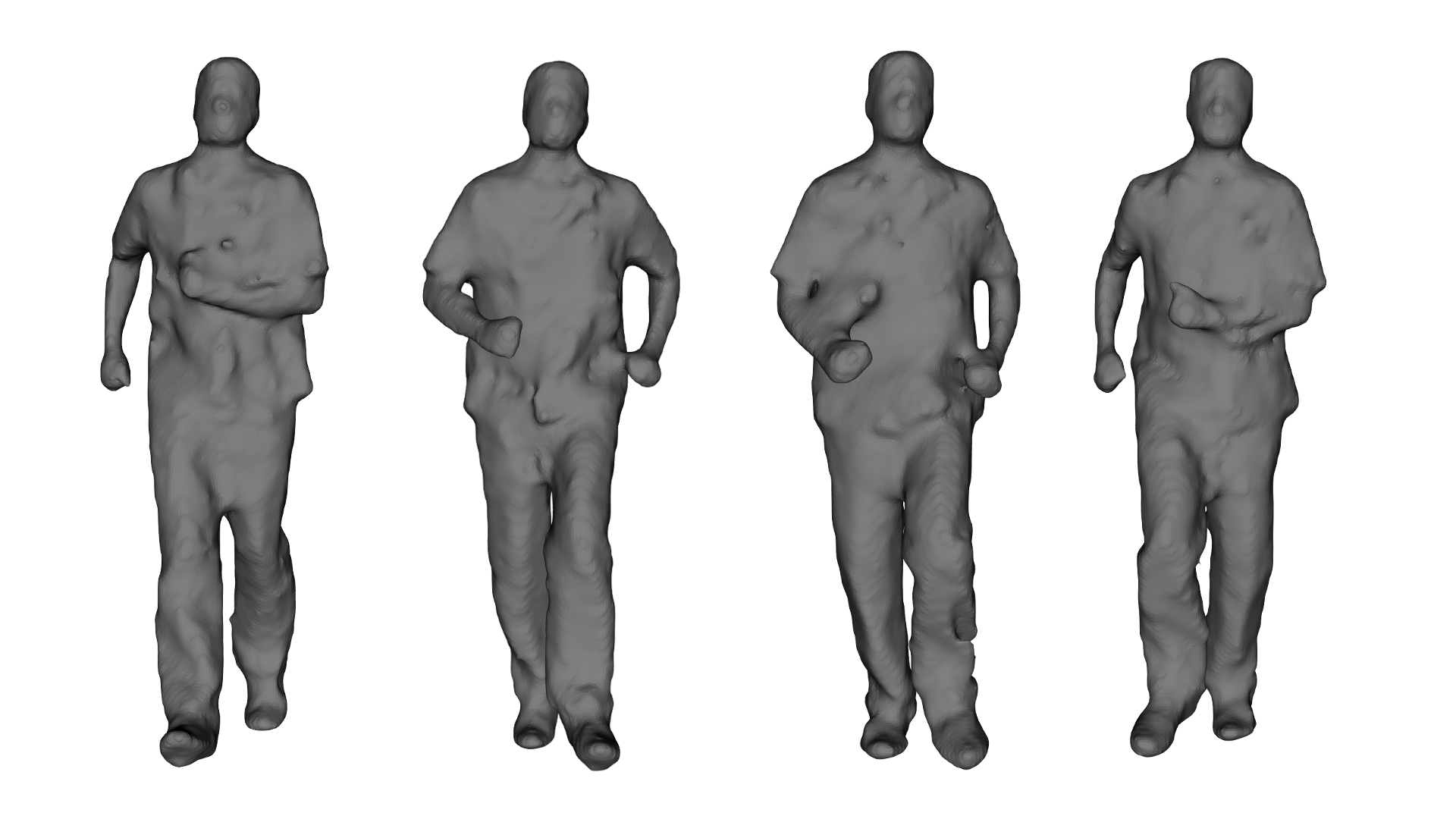}
    \end{minipage}
    
    \\ \hline 
  \end{tabular}
  \caption{First seven results of a query on the CVSSP3D real dataset using the Q-shape invariant as our representation.}
  \label{fig:query}
\end{table}

%%%%%%%%%%%%%%%%%%%%%%%%%%%%%
\subsection{Computation times}\label{sec:computation_times}

Our methods were implemented using Numpy routines, with no other optimization. The computations were performed with NumPy routines on a Intel(R) Core(TM) i5-7600K 3.8GHz CPU, with 8GB of RAM available.
% \textcolor{red}{In Table~\ref{tab:comp_nude} we present the mean computation time for the polygonal curve
% associated to a human motion when we use the breadths and areas feature vectors.  In Table~\ref{tab:time_harm} we present the computation time of 
% the polygonal curves when using the area spectrum, the breadth spectrum, and the shape 
% invariant. }

In Table~\ref{tab:faust_ctimes} we present the computation of each method. For Zhou \etal ~\cite{zhou20unsupervised} and Aumentado-Armstrong \etal~\cite{gdvae_2019}, we used the implementation, provided by the authors. For SMPL, we used the SMPL fitting pipeline proposed by the authors. 
In Table~\ref{tab:comp_cvssp} we present the computation time of each method for the CVSSP3D datasets. For Zhou \etal ~\cite{zhou20unsupervised} and Aumentado-Armstrong \etal~\cite{gdvae_2019} (GDVAE) we used the implementation provided by the authors. For Shape Distribution we use the hybrid Python-C implementation provided by Nenad Markuš~\footnote{\url{https://nenadmarkus.com/p/shape-distributions}}. For Spin Images, we used the C++ implementation provided by the PointCloud library~\footnote{\url{https://pointclouds.org/documentation/classpcl_1_1_spin_image_estimation.html}}.
We can see that our approach is the fastest on FAUST and CVSSP3D artificial datasets. We observe that the Q-shape invariant computation time is a bit slower than Shape Distribution for our approach in the real dataset -- but the performance of our approach improves the NN criteria by 5\%.

\begin{table}
    \centering
    \begin{tabular}{l|c|c}
    \hline
    Method & Real, 37800 vert. & Artif., 1290 vert. \\
    \hline
    \hline
    Shape Dist. & 79.1s* & 61.2s* \\ \hline
    Spin Image & 3h54* & 35.7s* \\ \hline
    %Zhou \etal & $/$ & 0.32s\\ \hline (no training)
    GDVAE & 56.4s & 2.08s \\\hline
    %Areas & 1.3s & 0.84 s \\ \hline
    %Breadths & 0.13 s & 0.095 s \\ \hline
    Shape invariant $\mathcal{E}^s_7$ & \textbf{46s} & \textbf{1.7s} \\ \hline
    Q-shape invariant & 209s &/ \\ \hline
    \end{tabular} 
    \caption{Mean computation time of polygonal curves extraction for different methods in the CVSSP3D datasets, along with the time corresponding to the polygonal curves in $\R^{24}$ using the Shape invariant $\mathcal{E}^s_7$, and the Q-shape invariant. We put the number of vertices for each dataset. Methods with an asterisk means that the implementation is not the official implementation provided by the authors.}
    \label{tab:comp_cvssp}
\end{table}

%\end{figure}
%%% Moved comparison of termporal alignment to after the end-document in main file.
\section{Conclusion and Future Work}
\label{sec:conclusion}
\subsection{Conclusion}
We defined a novel human descriptors using purely geometric information. Our approach is based on the intuition that a human pose is nearly characterized by its convex hull. Based on this hypothesis, we introduced three sequences of numerical surface descriptors that are invariant under reparametrizations, Euclidean transformations and similarities. We demonstrated the use of these descriptors by performing pose retrieval and extending their use to human motion retrieval. Our experiments on the FAUST and CVSSP3D synthetic and real datasets demonstrated that our 
method generally outperforms the state-art-methods for both 3D human pose and motion retrieval including deep learning approaches. \\
\subsection{Future Work}
Several avenues of future work are worth pursuing. We list some most promising directions below:
\begin{itemize}
    \item A first question is to ask if other descriptors~\cite{fourier_reconstruction, moments_reconstruction} of convex shapes with similar property as CH or EGI are suitable for describing the human pose.
    \item The noisy CVSSP3D real dataset has been a challenge for our descriptors. Some research should be spent on a statistical analysis as in in~\cite{poonawala_statistical_2002} to improve performance on noisy data.
    \item  As can be seen in Table~\ref{tab:cvssp_artif_NN}, the fusion of several descriptors does not automatically lead to better results. A finer statistical analysis is needed to exploit the existence of different descriptors. 
    \item It would be interesting to apply the geometric invariant and easily-computable descriptors proposed in this paper in a geometric deep learning approaches~\cite{bronstein2017geometric}.
\end{itemize}
\section{Acknowledgments}

This work was supported by the ANR project Human4D ANR-19-CE23-0020. This work was also partially supported by the French State, managed by National Agency for Research (ANR) under the Investments for the future program with reference ANR-16-IDEX-0004 ULNE.

%{\small
%\bibliographystyle{cag-num-names} 
%\bibliography{egbib}
%}

\end{document}